 \documentclass[preprint2]{aastex}

\usepackage{xspace}
\usepackage{pstricks}     
        \newsavebox{\IBox}
\usepackage{yhmath}
\usepackage{tikz}
\usetikzlibrary{shapes,arrows}

\usepackage{listings}

\usepackage{dcolumn}
\newcolumntype{d}[1]{D{.}{.}{#1} }

\usepackage{color}
    \definecolor{gray}{rgb}{0.5,0.5,0.5}
\usepackage{paralist} 

\newcommand{\myemail}{jdo@astro.unam.mx}

\def \ftest {$F$-test\xspace}
\def \ctest {$C$-test\xspace}
\def \aov {ANOVA\xspace}
\def \enhft {enhanced $F$-test\xspace}
\def \nest {nested ANOVA\xspace}
\def \rw {\ensuremath{\mathcal{N}}(0,0.006) RW\xspace}
\def \latgal {\ensuremath{b^{\mathrm{II}}}}
\def \qsoA {US\,3150\xspace}
\def \qsoB {1E\,15498+203\xspace}
\defcitealias{diego:2014}{Paper~I}

\slugcomment{Submitted to The Astronomical Journal}

\shortauthors{de Diego et al.}

\begin{document}


\title{Testing microvariability in quasar differential light curves using several field stars}

\author{Jos\'e A.\ de Diego\altaffilmark{1,2,3}, Jana Polednikova\altaffilmark{2,4}, Angel Bongiovanni\altaffilmark{2,4}, Ana M.\ P\'erez Garc\'{\i}a\altaffilmark{2,4}, Mario A. De Leo\altaffilmark{1,5}, Tom\'as Verdugo\altaffilmark{1} and Jordi Cepa\altaffilmark{2,4}}
\altaffiltext{1}{
	Instituto de Astronom\'{\i}a, 
	Universidad Nacional Aut\'onoma de M\'exico, 
    Avenida Universidad 3000,
    Ciudad Universitaria,
    C.P. 04510, Distrito Federal, Mexico \\
    \url{mailto:\myemail}}
\altaffiltext{2}{ 
    Instituto de Astrof\'{\i}sica de Canarias-Universidad de La Laguna}
\altaffiltext{3}{
    CEI Canarias: Campus Atl\'antico Tricontinental, 
    E-38205 La Laguna, Tenerife, Spain}
\altaffiltext{4}{
	Departamento de Astrof\'{\i}sica, Universidad de la Laguna, Spain}
\altaffiltext{5}{
	Department of Physics and Astronomy, 
    University of California, 
    Riverside, CA 92521, USA}


%



\begin{abstract}
%
%
\noindent
Microvariability consists in small time scale variations of low amplitude in the photometric light curves of quasars, and represents an important tool to investigate their inner core.
Detection of quasar microvariations is challenging for their non-periodicity, as well as the need for high monitoring frequency and high signal-to-noise ratio.
Statistical tests developed for the analysis of quasar differential light curves usually show either low power or low reliability, or both.
In this paper we compare two statistical procedures that include several stars to perform tests with enhanced power and high reliability.
We perform light curve simulations of variable quasars and non-variable stars, and analyze them with statistical procedures developed from the \ftest and the analysis of variance.
The results show a large improvement in the power of both statistical probes, and a larger reliability, when several stars are included in the analysis.
The results from the simulations agree with those obtained from observations of real quasars.
The high power and high reliability of the tests discussed in this paper improve the results that can be obtained from short and long time scale variability studies.
These techniques are not limited to quasar variability; on the contrary, they can be easily implemented to other sources such as variable stars.
Their applications to future research and to the analysis of large field photometric monitoring archives can reveal new variable sources.
\end{abstract}


\section{Introduction}



Variability is an important tool to study the inner physics and structure of the Active Galactic Nuclei (AGNs). 
The discovery of the variable behavior in the quasars \citep{matthews:1963} closely followed the discovery of quasars themselves. 
The time scales of the variations in the optical wavelenghts range from minutes to years. 
In general, the amplitude of the variation is larger when observed on longer time scales and shorter wavelengths. 
Large amplitude and time scale variations are relatively easy to detect, but the detection becomes problematic for small amplitudes and time scales of minutes.
Therefore, most if not all the microvariability studies are performed with telescopes of apertures larger than 1\,m \citep[e.g.,][]{kidger:1990, carini:1992, gopal:1995, ramirez:2004}.

We define optical microvariability as flux changes on time scales ranging from minutes to hours. The expected change in brightness for such events is in the order of hundredths of a magnitude \citep[e.g.][]{diego:1998}, posing a challenge for detection because the amplitude of the variation event and the noise level are similar. 
Moreover, quasar variability is aperiodic, and therefore difficult to analyze because we cannot collect observations to obtain a smoothly folded light curve.
These limitations place a strong emphasis on careful handling of the data and the use of robust and powerful statistical techniques for the analysis. 
Analyzing aperiodic variations has been the subject of recent work on AGNs \citep[the latter from now on \citetalias{diego:2014}]{emmanoulopoulos:2010, diego:2010, diego:2014}, as well as young stars and massive stars \citep{findeisen:2015}.

This issue was already recognized by several groups, that proposed various statistical approaches to search for microvariability events. 
Thus, there are several tests that have been frequently used to report microvariations. 
The \ctest as proposed by \citet{jang:1997} and \citet{romero:1999} is a very simple methodology that gained popularity ten years ago, but unfortunately it is not a valid statistical test \citep[\citetalias{diego:2014}]{diego:2010}. 
The \ftest is also a simple statistical procedure and has been replacing the \ctest during the last years.
However, the original version of the \ftest requires the quasar to be compared with a star of the same brightness, which is a condition that is difficult to meet. 
\citet{howell:1988} and \citet{joshi:2011} proposed the scaling of the photometric errors in order to work out this problem. 
Moreover, the \ftest faces another limitations: a lower power should be expected because of the violation of normality in data obtained from variable sources \citepalias{diego:2014}.
The one-way analysis of variance (\aov) was proposed by \citet{diego:1998} for microvariability detection, and it has been shown to be superior to the \ftest for this purpose \citep[\citetalias{diego:2014}]{diego:2010}.

Another strategy that has been considered is the use of multiple tests (multitesting) to face the problem of the low reliability of some studies, which can be affected by the odd behavior of a comparison star rather than the variability of the target quasar.
Multitesting has been implemented using two different tests or two different comparison stars \citep{joshi:2011, hu:2013}.
Microvariability detection is then claimed only if both tests, at the significance level of $\alpha = 0.01$, agree in the rejection of the null non-variability hypothesis.
However, in \citetalias{diego:2014} it was shown that this is a low power methodology that yields unreliable results.
Nevertheless, there is the possibility of including several comparison stars in the light curve analysis in such a way that the power of the test would be increased, as well as the reliability.
In this sense, the \enhft, proposed in \citetalias{diego:2014} involves several comparison field stars in a single \ftest, which provides a reliable methodology to increase the statistical power.

In this paper we present the \nest, which is an improved version of ANOVA that also includes several stars in the quasar differential light curve analysis. 
We compare \enhft and \nest results using both simulations and examples from available real data.

The paper is organized as follows. Section 2 describes the light curve simulations, the \enhft and the \nest methodology. In Section 3 we show and compare the results of the analysis of the simulated light curves, and the results obtained from real observations of the quasars \qsoA and \qsoB. Section 4 presents a brief discussion and conclusions.

%


\begin{figure}[t]
\centering
\includegraphics[trim = 0mm 0mm 0mm 20mm, clip, width=\linewidth]{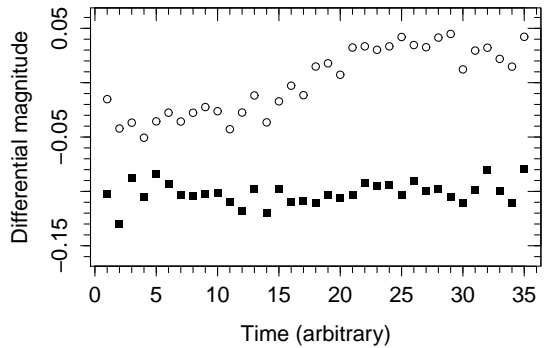}
\caption{Differential light curve examples.
Open circles show an example of a simulation of a variable quasar.
Filled rectangles show an example of a simulation of a control object.}\label{fig:lc}
\end{figure}

\section{Methods}\label{sec:methods}


This study consists of one thousand light curve simulations for a variable quasar of magnitude $V = 17$, a non-variable control object of the same brightness, and 21 field bright stars between magnitudes $16 \leq V \leq 18$, along with the comparison of the simulations with real observations. 
Figure~\ref{fig:lc} shows an example of the simulations of differential light curves: a variable quasar and a non-variable control star with the same photometric error.
Each differential light curve consisted of 35 observations obtained by subtracting the raw magnitudes of a reference star from the target to eliminate the changing-conditions effects between exposures.
This reference star is the brightest, non-saturated star in the field.

In contrast, the \aov procedures use only reference stars, but the test is performed between different groups of differential photometry observations of the same target rather than between the target and a comparison star light curves. 
As in \citetalias{diego:2014}, in the case of the variable quasars, the light curves were the result of a random walk function with steps normally distributed with mean 0 and standard deviation 0.006 [\rw\!\!], and a normally distributed photometric error of $\varepsilon \simeq 0.007$. 
The photometric errors for the set of 21 field stars are also normally distributed.
All the photometric errors correspond to 60\,s exposures in the $V$ band with the Harold L. Johnson 1.5\,m telescope at the Observatorio Astron\'omico Nacional in Sierra San Pedro M\'artir (M\'exico), and they were calculated using the simulator for this instrument.\footnote{Developed by Alan Watson, V. 1.2.0: \\ \url{http://www.astrossp.unam.mx/\%7Eresast/watsonBCh/\\fotometria-ccd.html}}
 
As detectors become larger and detector arrays more popular, there is a tendency towards larger field sizes and thus images that include more bright stars.
This situation makes it necessary to perform realistic simulations.
To deal with this necessity, the magnitudes for the 21 field stars were obtained using the star distribution in the $V$ band for a galactic latitude $\latgal = 60\degr$ \citep[Table 19.11, pp. 482-3]{allen:2000}.
The star distribution in magnitudes was approximated using a second order polynomial:
\begin{equation}
	log N_V = -4.026 + 0.5607 V - 0.00984 V^2,
\end{equation}
where $N_V$ is the cumulative function of the number of stars brighter than magnitude $V$ in one square degree of the sky.
The $V$ magnitudes for the 21 reference and comparison stars were obtained applying the inversion method of the cumulative function to the uniform distribution on the $N_V$ values in the range $16 \leq V \leq 18$, i.e. up to one magnitude of difference with respect to the quasar.
This range of magnitudes ensures that the images of the stars are suitable for differential photometry with the quasar, i.e. images are not suffering either overexposure or underexposure.
The parameters for the stars, in the order in which they appear in the simulations, are shown in Table~\ref{tbl:strpar}.

\begin{table}[t!]
\centering
\begin{center}
\caption{Star parameters for simulations.\smallskip}\label{tbl:strpar}
\begin{tabular}{rrr||rrr}
\tableline\tableline 
Index & \multicolumn{1}{c}{$V$} & Error & Index & \multicolumn{1}{c}{$V$} & Error \\
\tableline 
 1 & 17.727 & 0.011 & 11 & 16.119 & 0.004 \\ 
 2 & 17.998 & 0.013 & 12 & 16.655 & 0.006 \\ 
 3 & 17.798 & 0.012 & 13 & 17.392 & 0.009 \\ 
 4 & 17.173 & 0.008 & 14 & 17.336 & 0.009 \\ 
 5 & 17.351 & 0.009 & 15 & 17.850 & 0.012 \\ 
 6 & 16.709 & 0.006 & 16 & 17.493 & 0.010 \\ 
 7 & 17.761 & 0.011 & 17 & 17.856 & 0.012 \\ 
 8 & 16.633 & 0.006 & 18 & 17.077 & 0.007 \\ 
 9 & 16.185 & 0.004 & 19 & 17.703 & 0.011 \\ 
10 & 16.942 & 0.007 & 20 & 16.871 & 0.007 \\ 
   &        &       & 21 & 17.064 & 0.007 \\
\tableline
\end{tabular}
\end{center}
\end{table}

\subsection{Enhanced \ftest}

The simulations were analyzed using the \enhft, described in \citetalias{diego:2014}.
Until now, the \ftest has been implemented with a single comparison star, whose differential light curve is compared with the differential light curve obtained from a target object using the ratio of their respective variances.
It is important that the mean brightness of both the comparison star and the target object are matched to ensure that the photometric errors are equal, or at least that these errors are corrected to account for the differences in brightness, as proposed by \citet{howell:1988} and \citet{joshi:2011}.
The \enhft makes use of several comparisons stars, and it consists in transforming the comparison star differential light curves to have the same photometric noise as if their magnitudes matched exactly the mean magnitude of the quasar under study.
Note that the \enhft needs to use a reference star in order to obtain the differential light curves. 
Thus the number of available bright stars to perform the test is reduced by one, and the test is performed using only the rest of bright comparison stars.

\begin{figure*}[ht!]
\savebox{\IBox}{\includegraphics[trim = 0mm 0mm 0mm 19mm, clip, 
            width=0.5\linewidth]{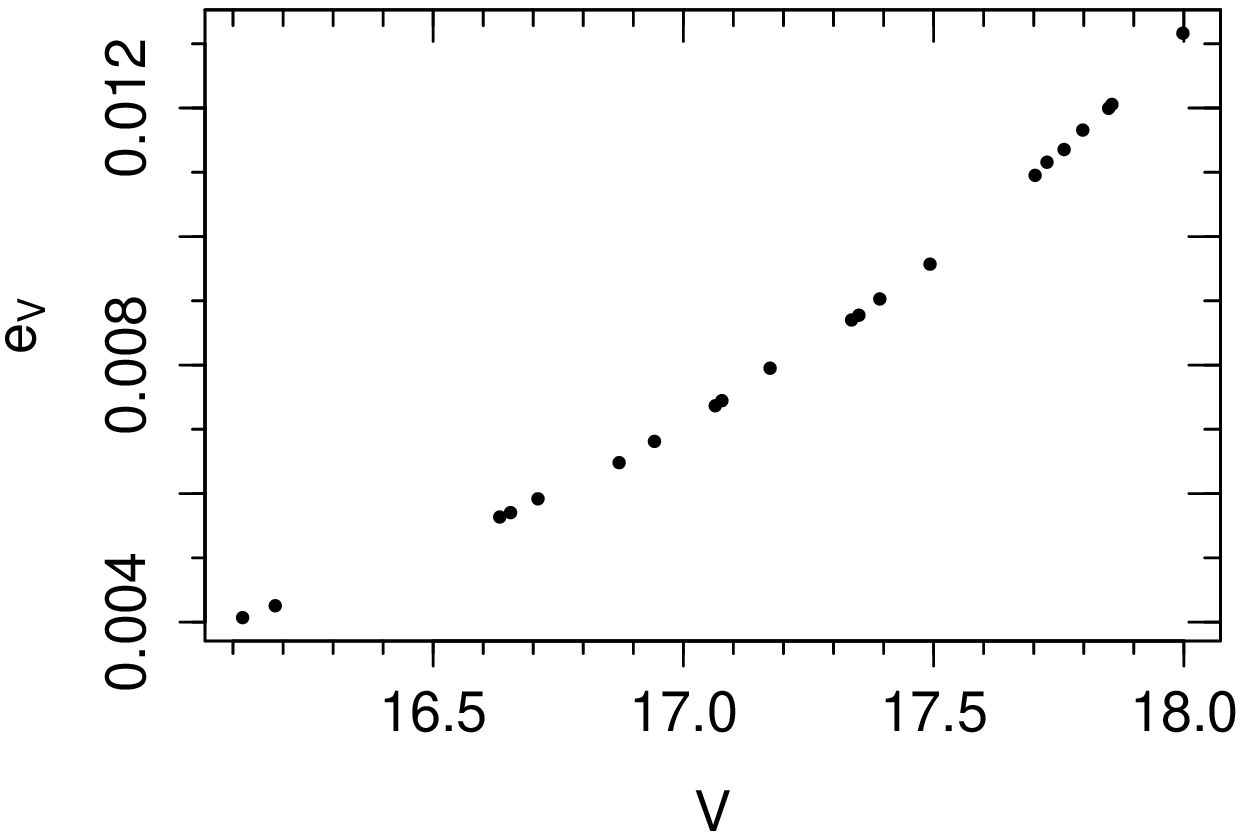}
        \includegraphics[trim = 0mm 0mm 0mm 19mm, clip, 
            width=0.5\linewidth]{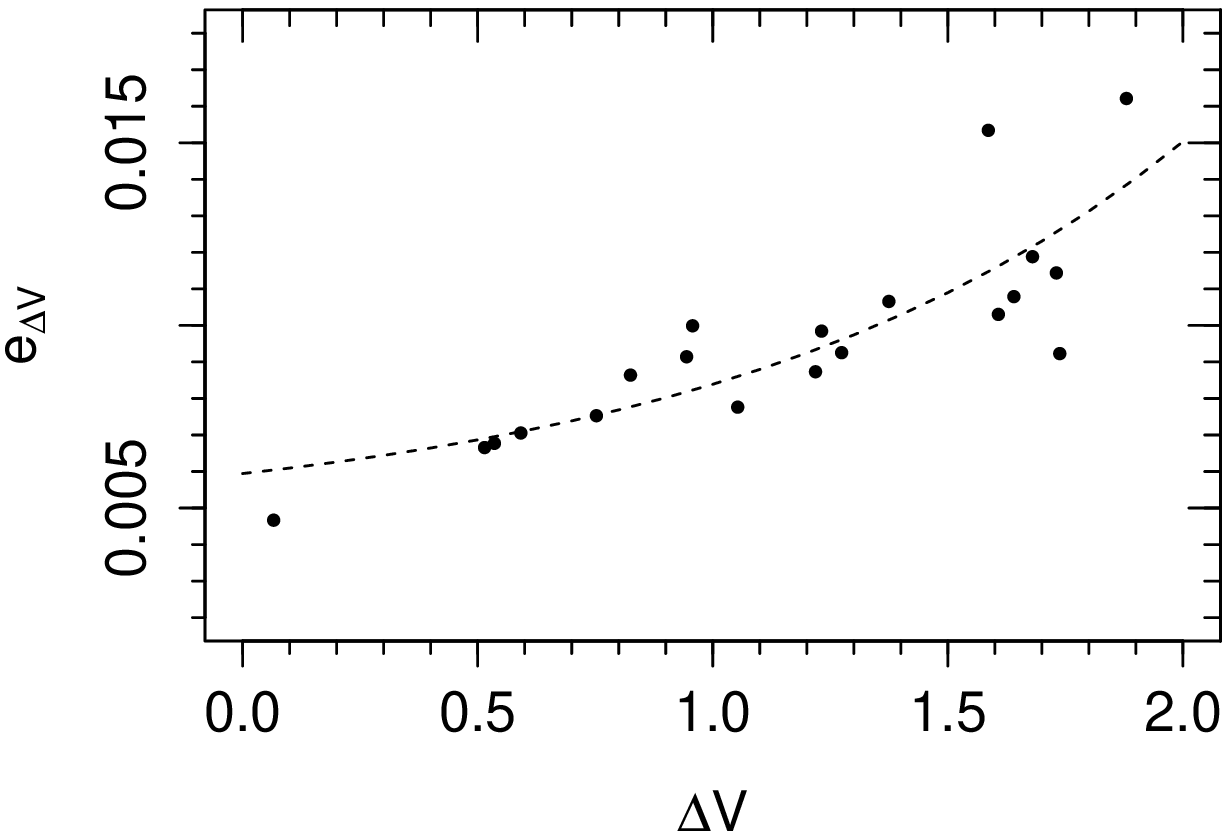}}
\begin{pspicture}(\wd\IBox,\ht\IBox)%
  \rput[lb](0,0){\usebox{\IBox}}%
  \rput[lt](.09\wd\IBox,.92\ht\IBox){(a)} 
  \rput[lt](.60\wd\IBox,.92\ht\IBox){(b)} 
\end{pspicture}
\savebox{\IBox}{\includegraphics[trim = 0mm 0mm 0mm 0mm, clip, 
            width=0.5\linewidth]{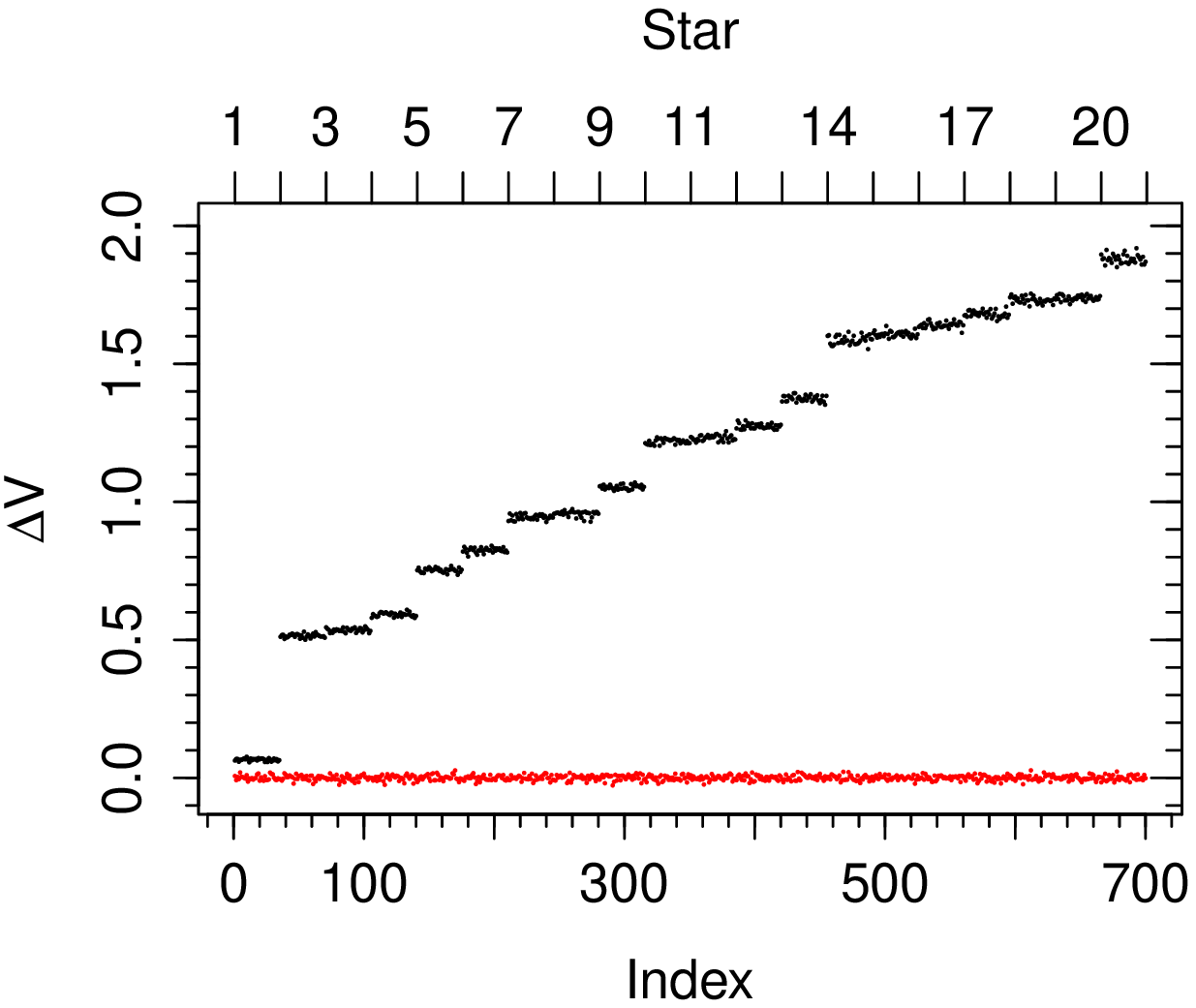}
        \includegraphics[trim = 0mm 0mm 0mm 0mm, clip, 
            width=0.5\linewidth]{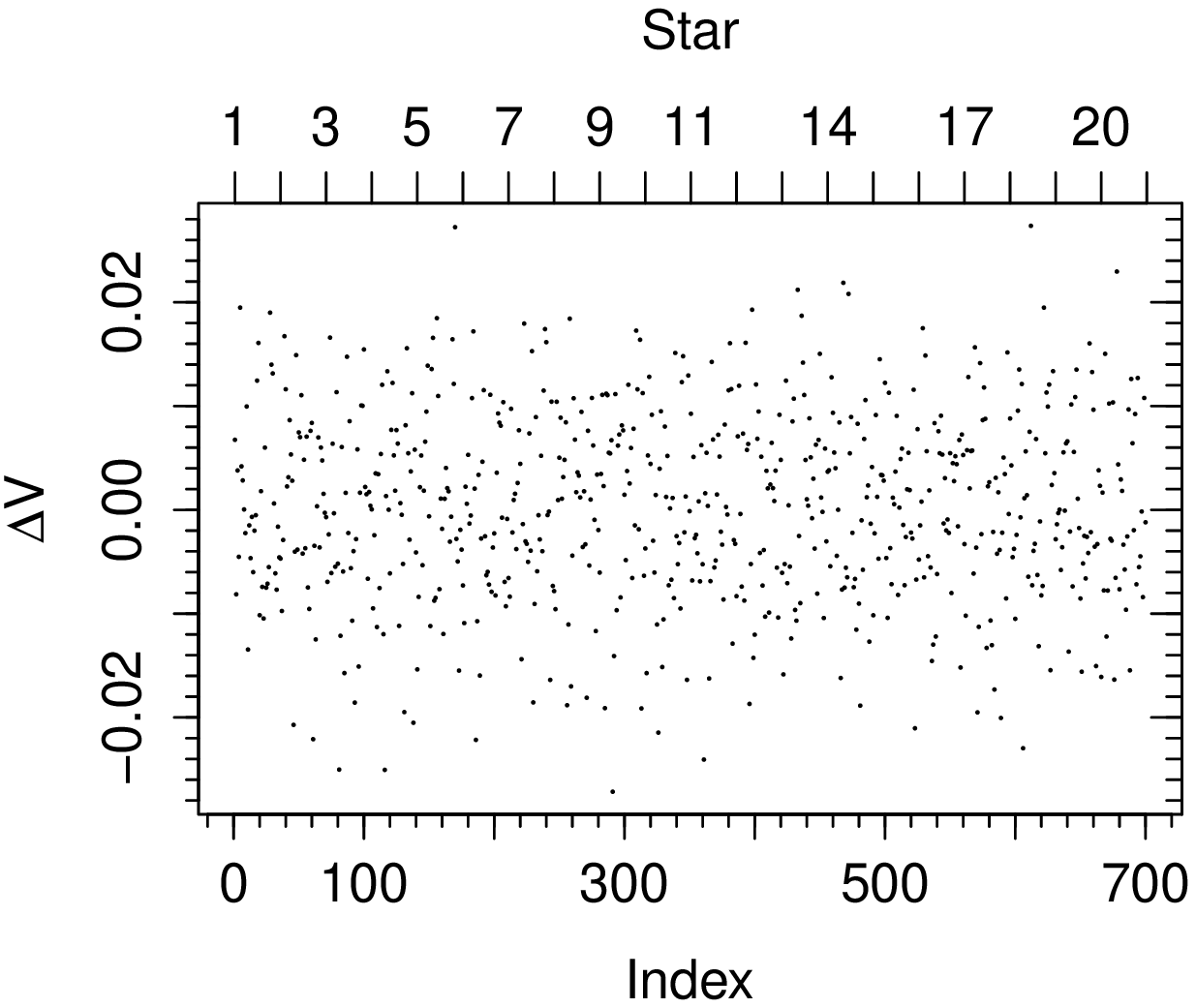}}
\begin{pspicture}(\wd\IBox,\ht\IBox)%
  \rput[lb](0,0){\usebox{\IBox}}%
  \rput[lt](.09\wd\IBox,.77\ht\IBox){(c)} 
  \rput[lt](.60\wd\IBox,.77\ht\IBox){(d)} 
\end{pspicture}
\caption{
	Enhanced \ftest steps. 
	Panel (a) shows the magnitudes of the 21 simulated stars and their associate errors obtained with the San Pedro Martir instrument simulator. 
	Panel (b) shows the differential light curve means and their associated errors for 20 comparison stars using the brightest star as reference; the dashed line indicates the exponential fit to the errors.
	Panel (c) shows the differential light curves for the 20 stars combined in a single set of $20\times35=700$ indexed observations (the star bins, of 35 observations each, are identified in the upper axis);after subtracting each star mean and transforming the errors to the same level as the quasar (in our case using the exponential fitting function), the 20 differential transformed light curves are stacked (lower red light curve).
	Panel (d) zooms on the stacked light curve shown in the previous panel.
 }\label{fig:enhftest}
\end{figure*}

Figure~\ref{fig:enhftest} helps us to understand the algorithm to perform the \enhft both in our simulations and in a real case. 
Figure~\ref{fig:enhftest}a shows the error and the $V$ magnitudes of the stars in a given field; the errors were obtained from the San Pedro Martir instrument simulator for the 1.5\,m telescope.
As in a real situation, we only need instrumental magnitudes to build the differential light curves.
We use the brightest star as our reference star to minimize errors; in our simulations this star is labeled with the number 11 (Table~\ref{tbl:strpar}).
Figure~\ref{fig:enhftest}b shows the errors vs. differential magnitudes.
We obtained these errors and magnitudes empirically from the simulated differential light curves, using their standard deviations and means, respectively.
The dashed line shows a fit to the differential magnitudes and errors data. 
In this paper we have used an exponential fit (see Section~\ref{sec:results}).
It is not important how we describe the dependence between instrumental errors and magnitudes, or even if we use an empirical \citep{joshi:2011} or theoretical \citep{howell:1988} relationship, as long as the fits are reasonable.
The point that matters is that we can estimate the differential photometric error for our target quasar.
For a quasar of magnitude $V_Q = 17$ and the bright reference star 11 ($V_{11} = 16.119$), the differential magnitude is $\Delta V_Q = V_Q - V_{11} = 0.881$, and the exponential fit yields an error $e_{\Delta V} = 0.008$.
Now we can transform the comparison star differential light curves to test the quasar variations.
Figure~\ref{fig:enhftest}c shows the differential light curves for each comparison star.
We subtract the mean for the $i$-star from the respective light curve, and multiply the resulting dataset by the ratio of the quasar and the star fitted errors: $e_{\Delta V}(Q) / e_{\Delta V}(St_i)$.
Then we have transformed the original comparison star differential light curve to values that can be compared with the quasar data.
Stacking all the transformed comparison star light curves, we obtain a single dataset to perform the \enhft (longer light curve at the bottom of the plot).
Note that in the \ftest the variances, rather than the means, are compared; our stacked dataset has zero mean, but it does not affect our results (an arbitrary constant value, for example the quasar mean differential magnitude, may be added to this dataset for graphical or other cosmetic purposes).
Figure~\ref{fig:enhftest}d shows the stacked dataset with more detail.

\subsection{Nested \aov}

We also analyzed the simulations using the \nest probe. 
The tests of the \aov family require groups of replicated observations.
In our case, these groups are arranged by five observations of a given quasar in a time lapse of around 5 to 10 minutes for telescopes with apertures larger than 1\,m, so that the quasar brightness can be considered constant for the observations collected in the same group.
This is the same procedure used in previous real observations \citep{diego:1998, ramirez:2004, ramirez:2009} and simulations \citep[\citetalias{diego:2014}]{diego:2010}.
Of course, microvariations shorter than the lapse time within the group observations would not be detected by ANOVA, but such phenomena, known as spikes, has been seldom reported in the literature \citep{sagar:1996, diego:1998, gopal:2000, stalin:2004}.

The \nest, as implemented in this paper, consists in the study of the variances at three different experimental levels.
The first level is the one in which we are really interested: the differences between the groups, that are a signature of the quasar variability.
The second level corresponds to the differences between the observations due principally to shot noise and sky subtraction.
The third level accounts for the variance between the tests caused by the different reference stars.
A complete description of the \nest procedure is given in Appendix~\ref{app:nesttheor}.
Calculations have been performed using the R language, and the \nest probes were carried out using the \textbf{aov} function included in the \emph{stats} package.
The complete code used for the \nest analysis applied to real data can be found in Appendix~\ref{app:nestcode}.

Note that \aov family tests compare the dispersion of the individual differential magnitudes of the quasar within the groups, and the dispersion between the groups, without using any comparison star.
Instead, all the bright stars are utilized as reference stars to build distinct differential light curves, and thus we will always be able to include one more star in the \nest analysis than in the \enhft.


\section{Results}\label{sec:results}


\subsection{Simulations}

The results in \citetalias{diego:2014} showed that \aov has more power than the \ftest to detect variability when only one comparison star is taken into account in the analysis.
Moreover, the power of the \ftest was reduced by the probably non-normal distribution of the variable quasar data.
This power loss can be compensated by using the \enhft, that includes more comparison stars in the analysis.
Here we present the results of increasing the number of stars using \nest that lead to a test power improvement, and we compare the outcomes obtained using both the \nest and the enhanced $F$ tests.

\begin{figure*}[ht]
\savebox{\IBox}{\includegraphics[trim = 0mm 21mm 0mm 17mm, clip, 
            width=0.5\linewidth]{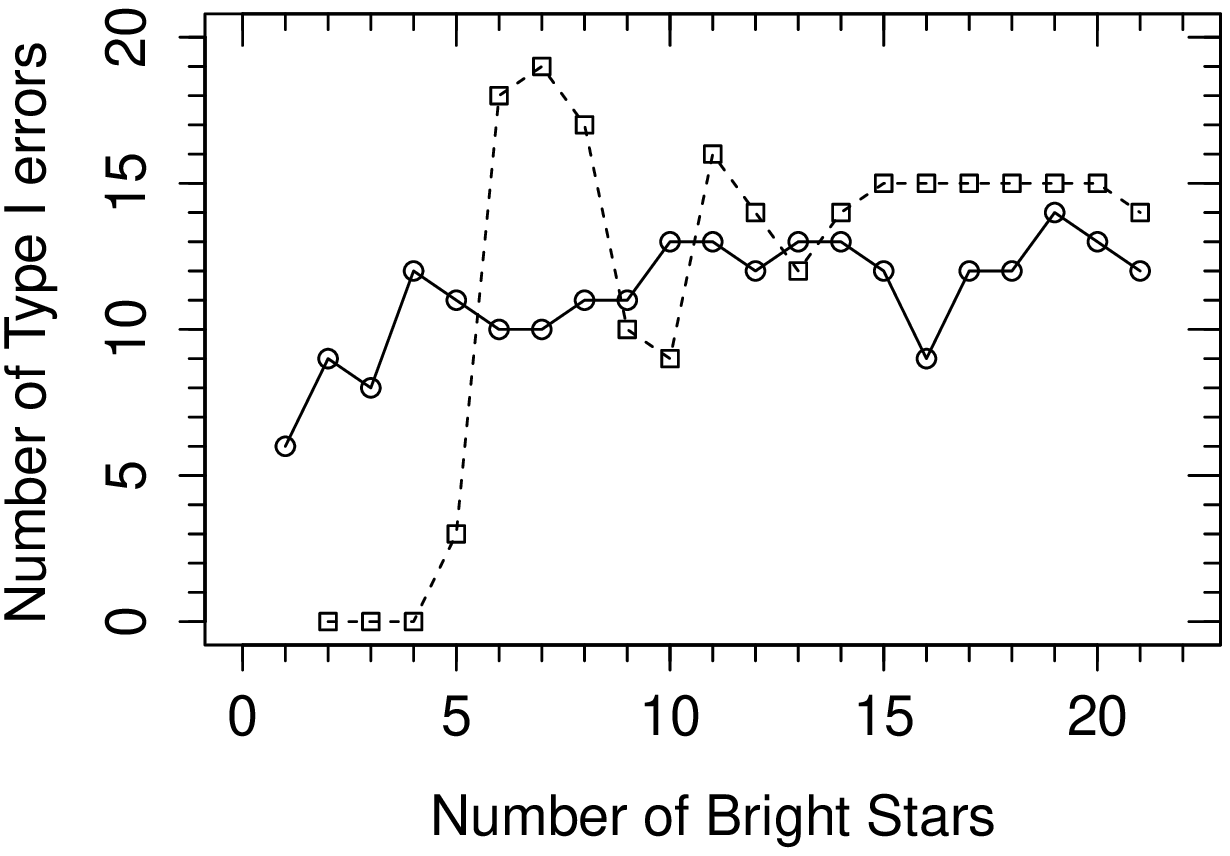}
        \includegraphics[trim = 0mm 21mm 0mm 17mm, clip, 
            width=0.5\linewidth]{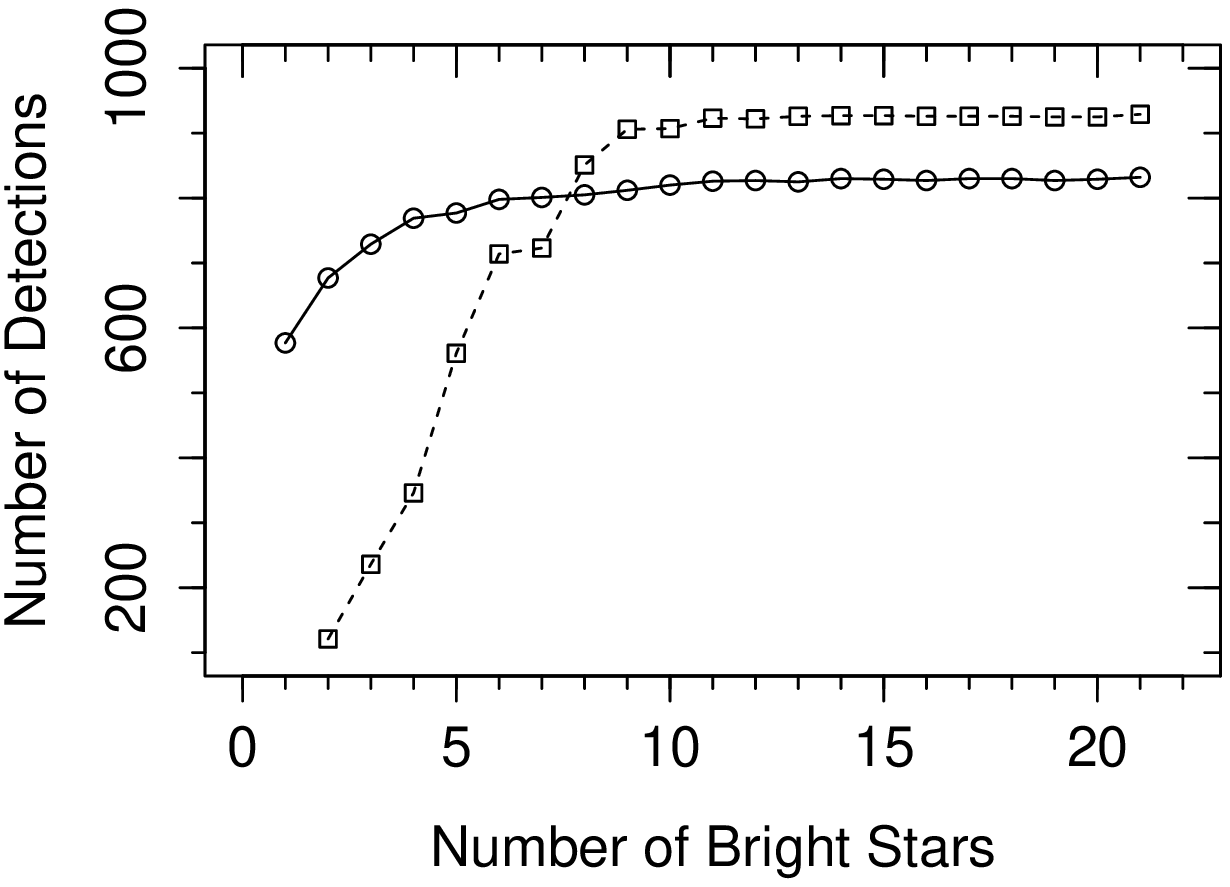}}
\begin{pspicture}(\wd\IBox,\ht\IBox)%
  \rput[lb](0,0){\usebox{\IBox}}%
  \rput[lt](.09\wd\IBox,.85\ht\IBox){(a)} 
  \rput[lt](.59\wd\IBox,.85\ht\IBox){(b)} 
\end{pspicture}
\savebox{\IBox}{\includegraphics[trim = 0mm 0mm 0mm 19mm, clip, 
            width=0.5\linewidth]{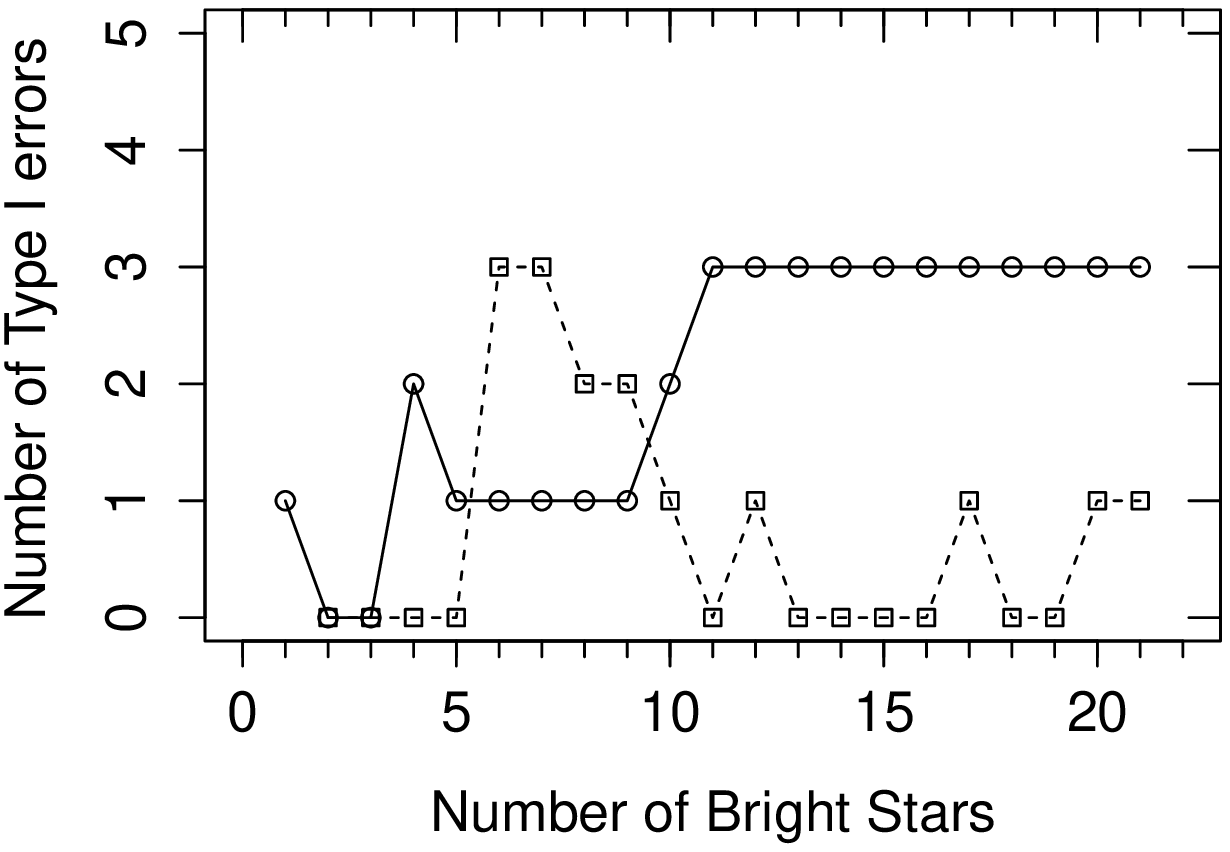}
        \includegraphics[trim = 0mm 0mm 0mm 19mm, clip, 
            width=0.5\linewidth]{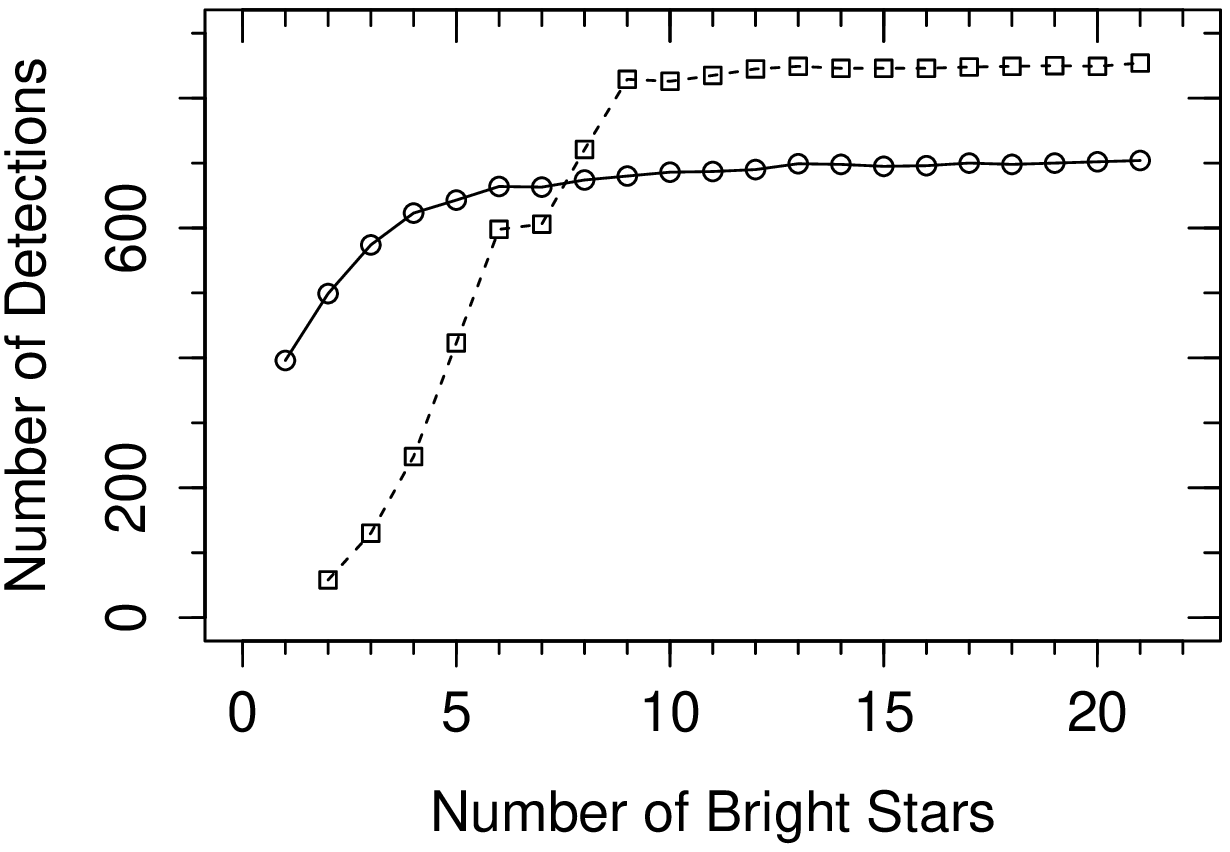}}
\begin{pspicture}(\wd\IBox,\ht\IBox)%
  \rput[lb](0,0){\usebox{\IBox}}%
  \rput[lt](.09\wd\IBox,.90\ht\IBox){(c)} 
  \rput[lt](.59\wd\IBox,.90\ht\IBox){(d)} 
\end{pspicture}
\caption{Test outcomes for 1000 light curve simulations. 
Results for the \enhft and \nest are shown as dashed line connected squares, and solid line connected circles, respectively.
Tests have been performed at the significance levels $\alpha = 0.01$ [panels (a) and (b)] and $\alpha = 0.001$ [panels (c) and (d)].
Type I errors obtained from the control object are shown in panels (a) and (c), and the number of null hypothesis rejections for the varying quasar are shown in panels (b) and (d).
The magnitudes and photometric errors for the stars have been obtained from a realistic star field at galactic latitude $\latgal = 60\degr$, as explained in the text.}\label{fig:realpower}
\end{figure*}

The stars in the simulations have different magnitudes and errors, as explained in Section~\ref{sec:methods}.
Figure~\ref{fig:realpower} shows the number of false detections or Type I errors, and the number of detections.
The stars are included in the same order as in Table~\ref{tbl:strpar} to reproduce the characteristics of the available objects as the field size increases (i.e., the less numerous bright stars will show up as the field size increases).
However, note that the final result for the test with a given number of stars is independent of the order in which these stars are introduced for both the \enhft and \nest.
The reason why the number of stars are different for both tests is that both the \ftest and the \enhft require an extra reference star for differential photometry and the other stars for comparison, while for the \nest test all the stars are used as reference (no comparison stars are necessary).
As a result, given a set of bright stars in the quasar field, we will always have one more star to perform the \nest probe than the \enhft.
To account for this effect, the niche for the bright star number 1 in Figure~\ref{fig:realpower} is empty for the \enhft, but not for the \nest.
The number of bright stars is directly proportional to the size of the field.
For a galactic latitude $\latgal = 60\degr$, we expect around 0.13 stars per square arcminute, i.e. around 2 stars in a $4\arcmin \times 4\arcmin$ detector. 

As the number of stars included in the analysis increases, the tests are more and more powerful tending to an upper asymptotic value. 
We see in Figure~\ref{fig:realpower} that in our simulations the number of detections for the \enhft is larger than the number for the \nest for a number of eight or more stars.
The \enhft starts with very few detections for a few and dim stars (that dominate the star distribution), but the number of detections increases dramatically and overcomes the detection for the \nest as the number of stars increases and bright stars are introduced into the analysis.

\begin{figure}[htp]
\savebox{\IBox}{\includegraphics[trim = 0mm 21mm 0mm 20mm, clip, width=\linewidth]{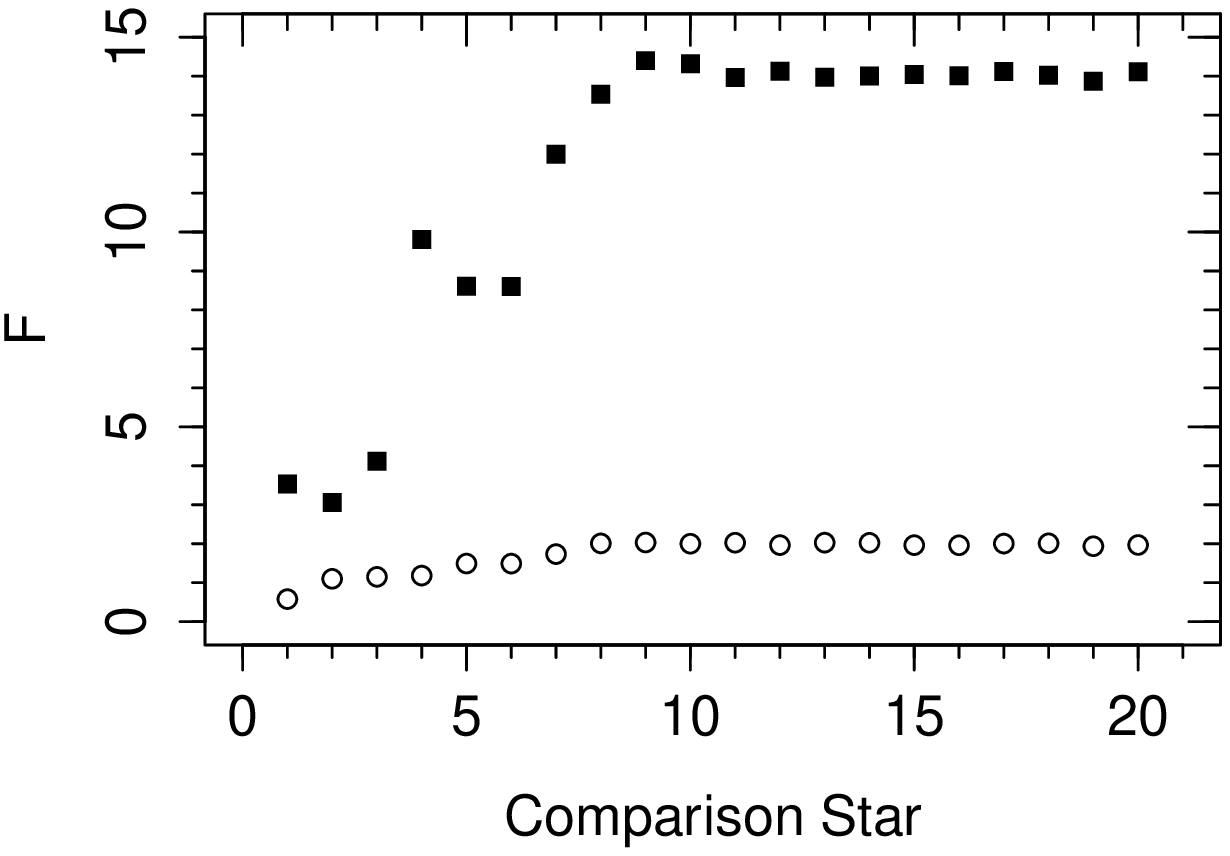}}
\begin{pspicture}(\wd\IBox,\ht\IBox)%
  \rput[lb](0,0){\usebox{\IBox}}%
  \rput[lt](.8\wd\IBox,.65\ht\IBox){(a)} 
\end{pspicture}
\savebox{\IBox}{\includegraphics[trim = 0mm 21mm 0mm 19mm, clip, width=\linewidth]{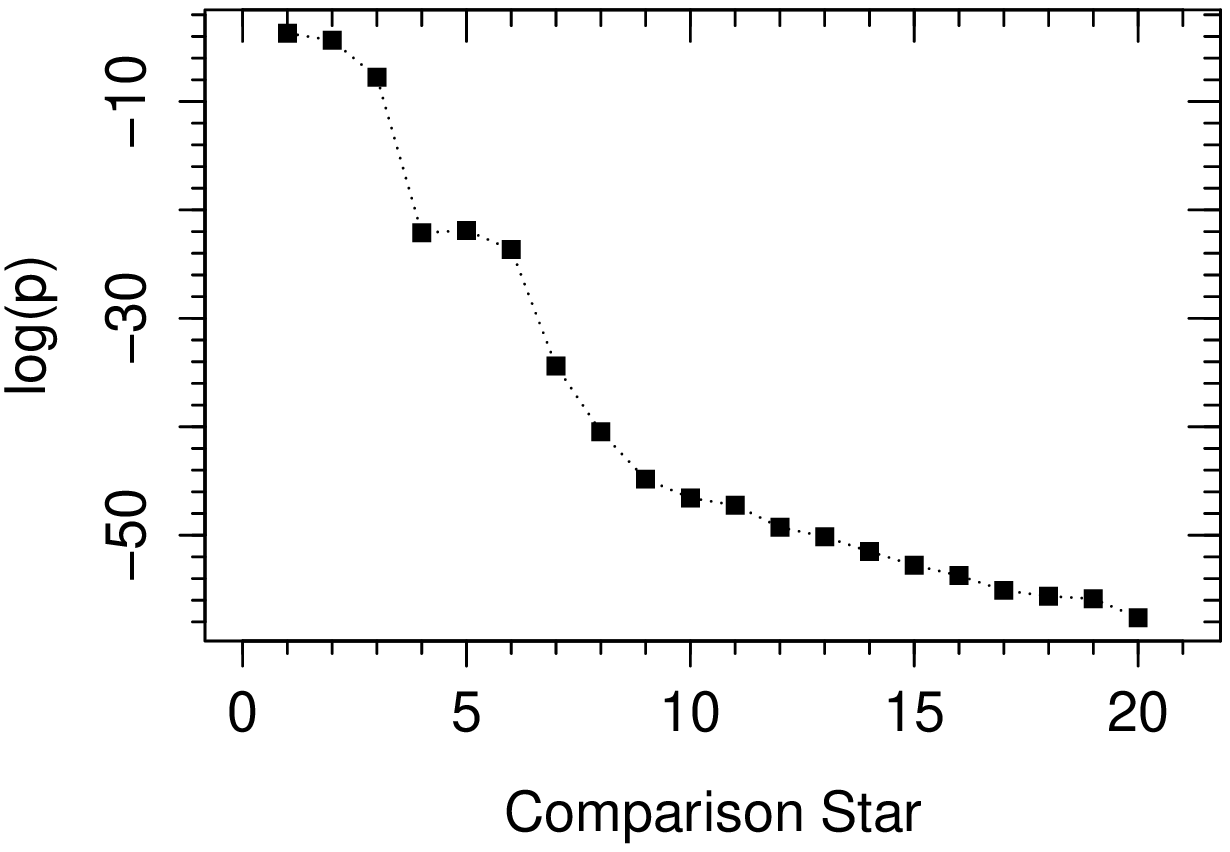}}
\begin{pspicture}(\wd\IBox,\ht\IBox)%
  \rput[lb](0,0){\usebox{\IBox}}%
  \rput[lt](.8\wd\IBox,.65\ht\IBox){(b)} 
\end{pspicture}
\savebox{\IBox}{\includegraphics[trim = 0mm 0mm 0mm 19mm, clip, width=\linewidth]{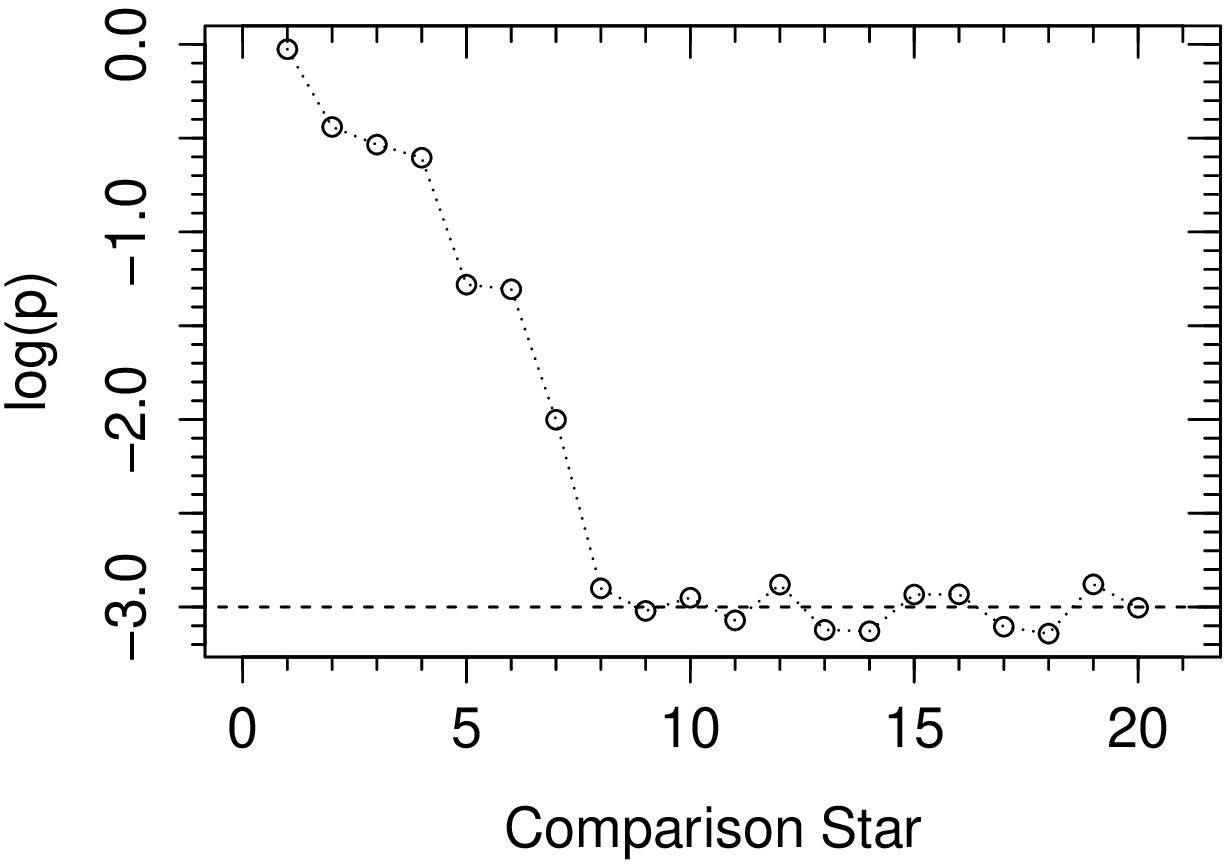}}
\begin{pspicture}(\wd\IBox,\ht\IBox)%
  \rput[lb](0,0){\usebox{\IBox}}%
  \rput[lt](.8\wd\IBox,.75\ht\IBox){(c)} 
\end{pspicture}
\caption{Two examples of the \enhft output.
    Panel (a) shows the $F$ value calculated as the variance ratio between the simulated quasar light curve and the stacked star light curves for a clearly variable quasar (filled squares) and a less variable one (hollow circles).
    Panel (b) shows the logarithm of the probabilities that the data from the earlier quasar have been drawn from a non-variable object.
    The probabilities show a decreasing trend, as expected from the power increase when more stars are added in the analysis.
    Panel (c) is like the panel (b) but for the later quasar.
    Although the $F$ values show very small variations for this object, and the probabilities show the expected trend as in panel (b), the tests are just in the limit of detection at the 0.001 level (dashed line), and tend to fluctuate between detections and non-detections.}\label{fig:lcstat}
\end{figure}

The behavior of both the \nest and the \enhft can be explained as the result of the test designs and the data attributes.
In the case of \nest, all the stars are used as reference stars, and all of them have the same weight in this probe.
However, the results of the \enhft depend very much on the photometric errors of the reference star.
These errors affect the precision of the whole differential photometry.
As the dimm stars dominates the star distribution, we need large fields to include bright stars.
We can see this effect on the \enhft detection curves in Figures~\ref{fig:realpower}b and \ref{fig:realpower}d.
When a bright star such as the star number 6 becomes the reference star, the number of the \enhft and \nest detections are comparable, and when the even brighter stars 8 and 9 successively became the reference stars (and star 6 becomes a comparison star), the \enhft overcomes \nest.

The inclusion of comparison stars brighter and dimmer than the quasar allows a better fit for the photometric errors as a function of the differential magnitudes.
If all the stars are dimmer than the quasar, the exponential fit that we have used must be extrapolated to the quasar brightness, which is less accurate than interpolating the error.
In fact, we have found that this inaccuracy produces too many type I errors.
This may be an inconvenience in the cases of few field stars or of bright quasars.
If comparison stars brighter than the quasar are not available, at least for simulated data it is safer to give more weight to the star with the brightness which is more similar to that of the quasar, although this procedure reduces the power of the \enhft.
In a real situation, the errors may be individually estimated using another procedure \citep[e.g.][]{howell:1988, joshi:2011}, but generalizing such a procedure for simulated data is beyond the scope of this paper.

\begin{figure*}[t]
\savebox{\IBox}{\includegraphics[trim = 0mm 21mm 9mm 17mm, clip, 
            width=0.5\linewidth]{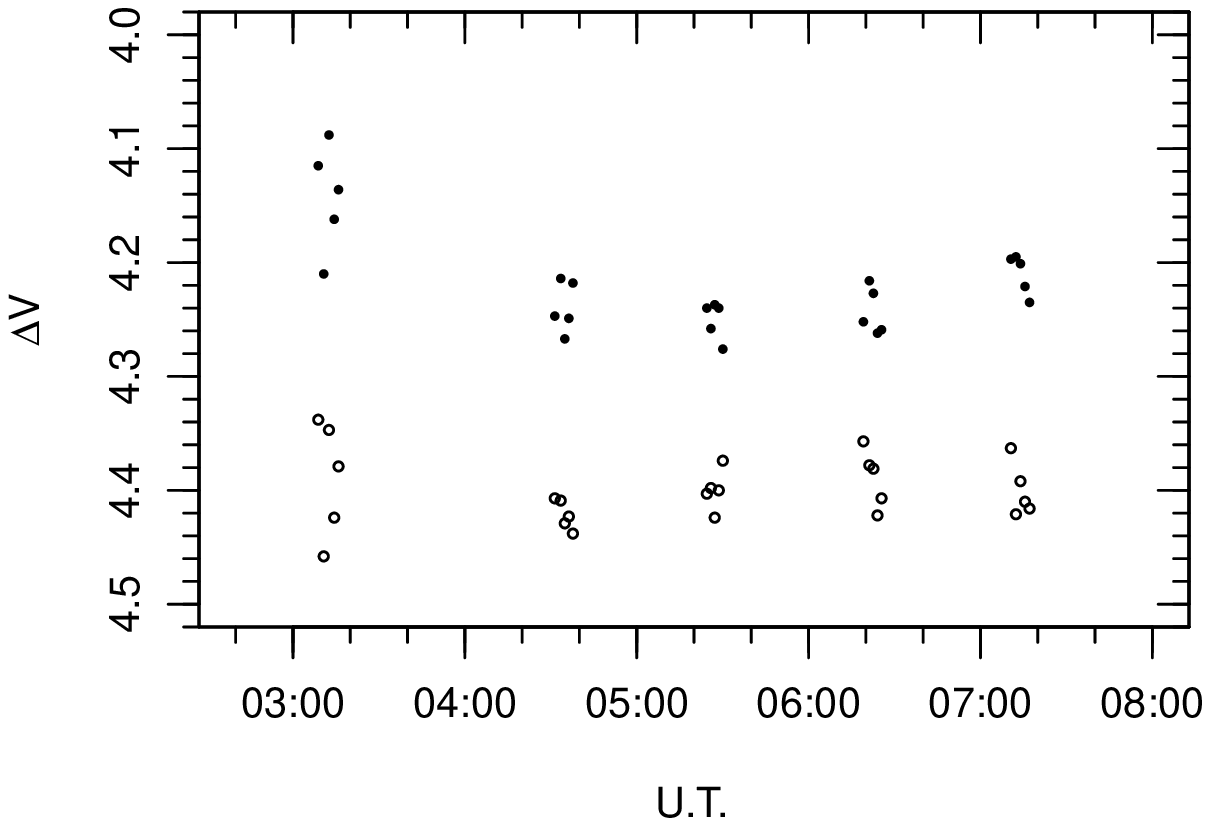}
        \includegraphics[trim = 0mm 21mm 9mm 17mm, clip, 
            width=0.5\linewidth]{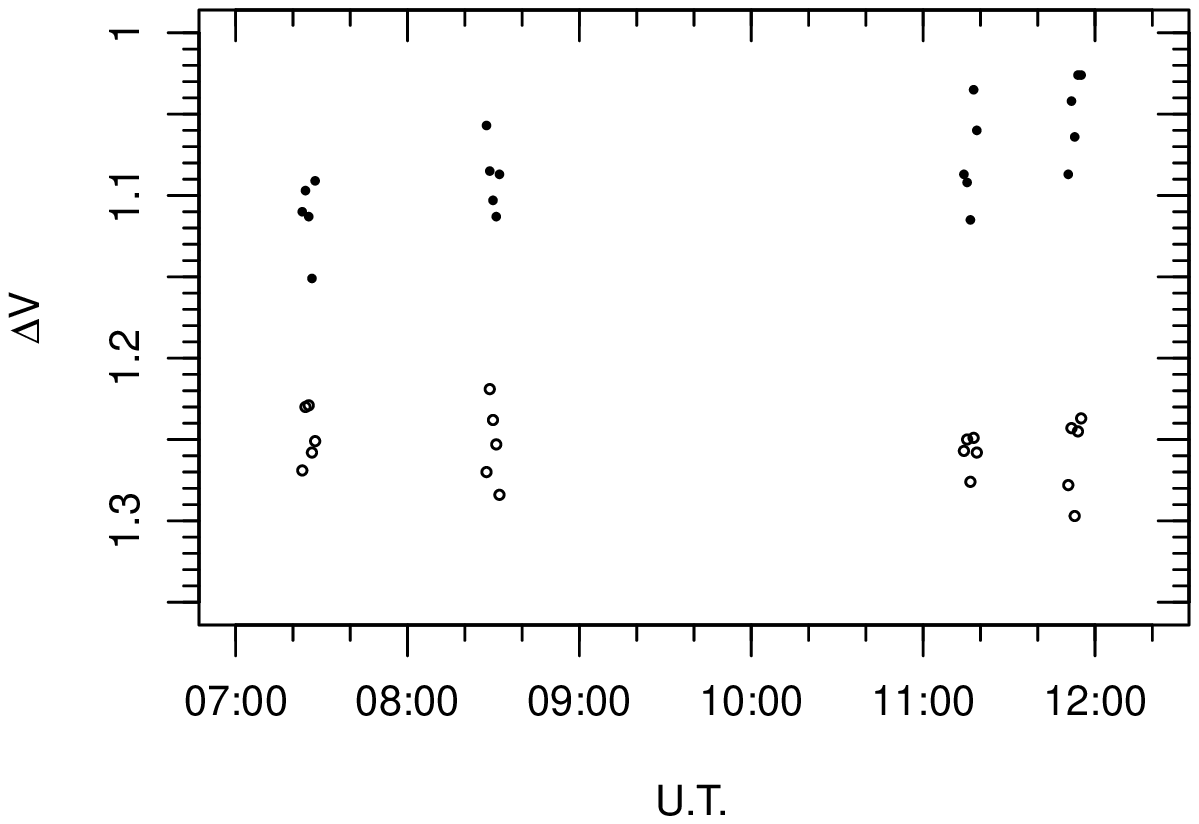}}
\begin{pspicture}(\wd\IBox,\ht\IBox)%
  \rput[lb](0,0){\usebox{\IBox}}%
  \rput[lt](.10\wd\IBox,.86\ht\IBox){(a)} 
  \rput[lt](.605\wd\IBox,.86\ht\IBox){(c)} 
\end{pspicture}
\savebox{\IBox}{\includegraphics[trim = 0mm 0mm 9mm 19mm, clip, 
            width=0.5\linewidth]{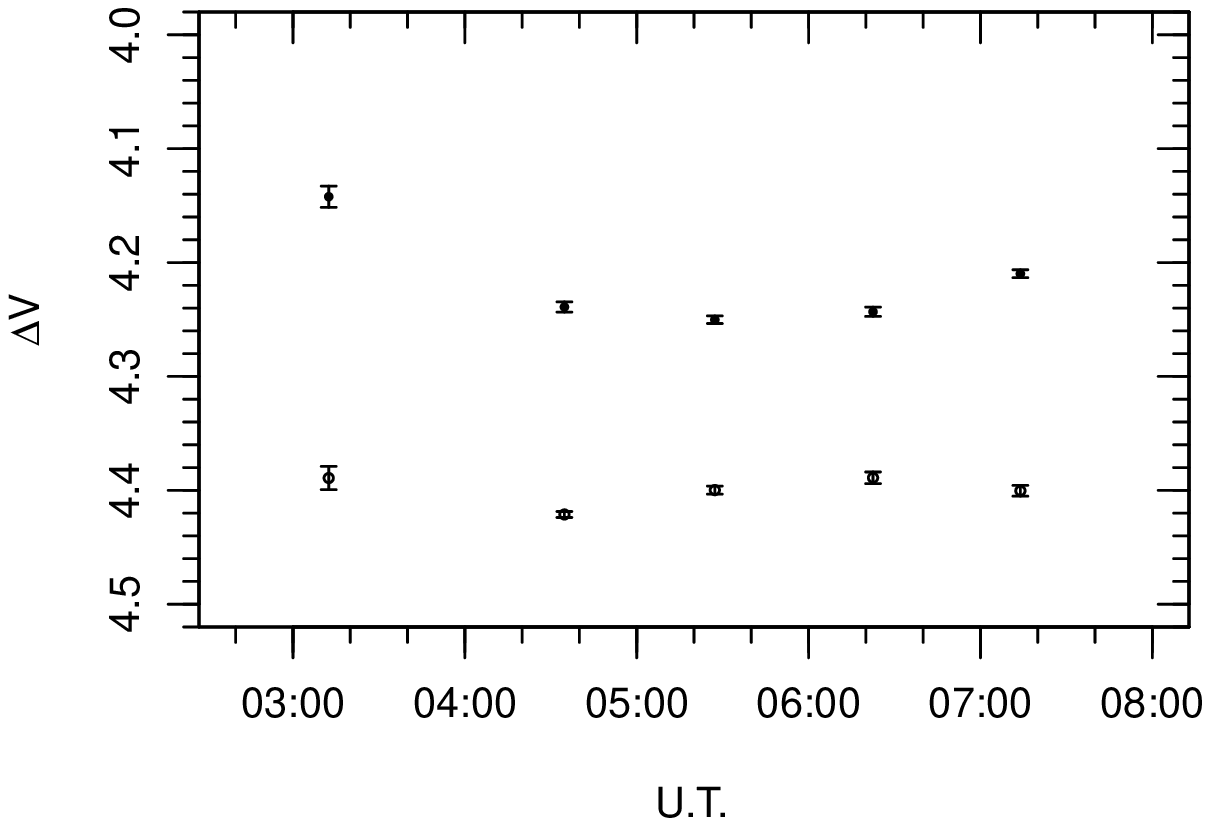}
        \includegraphics[trim = 0mm 0mm 9mm 19mm, clip, 
            width=0.5\linewidth]{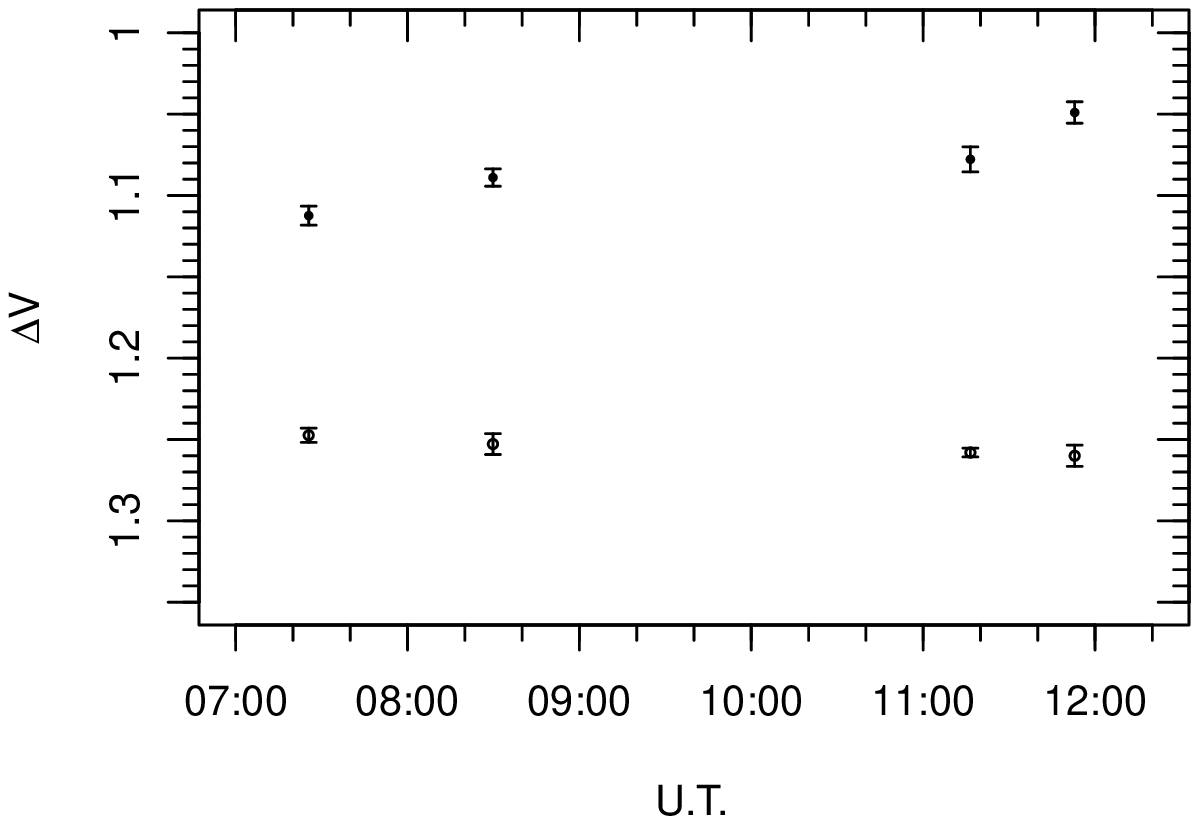}}
\begin{pspicture}(\wd\IBox,\ht\IBox)%
  \rput[lb](0,0){\usebox{\IBox}}%
  \rput[lt](.10\wd\IBox,.91\ht\IBox){(b)} 
  \rput[lt](.605\wd\IBox,.91\ht\IBox){(d)} 
\end{pspicture}
\caption{
	Monitored light curves for quasars \qsoA (November 13, 1996), and \qsoB (May 12, 1997) and a comparison star.
	Panels (a) and (b) show the \qsoA raw and binned differential light curves (upper points) and a comparison star (lower points).
	Error bars in panels (b) and (d) correspond to the standard error measured within each group of observations.
	}\label{fig:obslc}
\end{figure*}

As the tests with more stars are not fully independent of the test with less stars (for example, 14 of the 15 stars included in the 15 star tests are the same as in the tests for 14 stars), the Type I errors are not random, but show a \emph{memory} of the errors obtained in the previous tests.
Finally, the effect size estimate (for example, the $F$ value of the ratio of variances in the \enhft) is fairly constant when adding more stars in the analysis. 
Although $F$ approaches an asymptotic  value and the probabilities decline smoothly with the number of stars (Figure~\ref{fig:lcstat}b), for a given light curve the analysis may yield fluctuations in the probabilities obtained with different number of stars.
In those cases when the $F$ value is close to the critical limit of detection, the tests may fluctuate between detections and non-detections as shown in Figure~\ref{fig:lcstat}c.

\begin{table*}[t]
\centering
\begin{center}
\caption{Test results for \qsoA data.\smallskip}\label{tbl:us3150}
\begin{tabular}{ccc@{ }cccccc@{ }ccc}
\tableline\tableline \noalign{\smallskip}
 & \multicolumn{5}{c}{\qsoA} && \multicolumn{5}{c}{Control} \\
\cline{2-6} \cline{8-12} \noalign{\smallskip} 
 N. Stars & \multicolumn{2}{c}{\ftest} && \multicolumn{2}{c}{\aov} && \multicolumn{2}{c}{\ftest} && \multicolumn{2}{c}{\aov} \\
 \cline{2-3} \cline{5-6} \cline{8-9} \cline{11-12} \noalign{\smallskip} 
 & $F$ & $p$ && $F$ & $p$ && $F$ & $p$ && $F$ & $p$ \\
\tableline \noalign{\smallskip}
1 & 2.240 & 2.2E-2 && 13.44 & 1.8E-5 && 0.622 & 0.87 && 0.995 & 0.43 \\
2 & 2.573 & 2.9E-3 && 15.71 & 5.8E-6 && 0.671 & 0.85 && 1.244 & 0.33 \\
3 & 2.450 & 2.0E-3 && 17.52 & 2.6E-6 && 0.639 & 0.89 && 1.123 & 0.37 \\
4 & 2.525 & 8.1E-4 && 15.34 & 6.9E-6 &&  --   &  --  && 1.767 & 0.18 \\
5 &  --   &   --   && 17.06 & 3.1E-6 &&  --   &  --  &&  --   &  --  \\
\tableline
\end{tabular}
\tablecomments{The number of degrees of freedom for the \enhft is 24 for the numerator and 22, 46, 70 and 94 for the denominator, for 1, 2, 3, and 4 stars respectively.
For the \nest, the number of degrees of freedom is always 4 for the numerator and 20 for the denominator. }
\end{center}
\end{table*}

\subsection{Observations}

Data discussed in this section correspond to the radioquiet quasars \qsoA and \qsoB, and have been already reported in \citet{diego:1998}.
Here we present a new analysis including more field stars to compare the results obtained with \nest and the \enhft.
The quasar light curves are shown in Figure~\ref{fig:obslc}.
Observations of \qsoA ($V \simeq 16.8$) were obtained during about 6\,h of monitoring on November 13, 1996 with the $f/13.05$ 1.5\,m telescope at San Pedro Martir Observatory (Baja California, Mexico) equipped with a Thomson THX 31156 CCD of $1024 \times 1024$ pixels of $19 \times 19\ \mathrm{\mu m}$ and Metachrome II coated.
Quasar \qsoA showed clear microvariability at a level of significance of $\alpha = 0.001$, an amplitude of $\Delta m \simeq 0.1$\,mag.
The data consist of 7 groups of 5 observations with exposures of 60\,s in the $V$ band.
There were five stars in the $3\farcm4 \times 3\farcm4$ field suitable for our purpose, but at the time when we performed the observations, we had no plans to use all these stars in our analysis.
Hence, the brightest star (more than 3 magnitudes brighter than the quasar) was overexposed in the two last groups of observations.
Therefore, only for demonstration purposes, we have dropped these two groups and we performed the analysis using the first 5 sets of observations, corresponding to about 4\,h of monitoring.

Observations of \qsoB ($V \simeq 17.12$) were obtained during 4.5\,h of monitoring on May 12, 1997 with the same telescope, but this time equipped with a Tektronix TK1024AB of $1024 \times 1024$ pixels of $24 \times 24\ \mathrm{\mu m}$, thinned and Metachrome II coated.
Quasar \qsoB showed only weak evidence of variability at $\alpha = 0.01$, an amplitude of $\Delta m \simeq 0.06$\,mag.
The data consists of 5 groups of 5 observations with exposures of 60\,s in the $V$ band.
There were four stars in the $4\farcm3 \times 4\farcm3$ field suitable for our purpose, but photometry was affected by the scattered light from the bright star HD\,142109 ($V = 8.94$) at around 1\arcmin\ from the quasar, which spoils the observations of some stars in the third group.
Consequently, we have dropped this group from our analysis.

Tables~\ref{tbl:us3150} and \ref{tbl:1e15498} show the results obtained with the \enhft and \nest for \qsoA and \qsoB, respectively.
In both tables, column 1 indicates the number of stars used in the analysis.
Columns 2 and 3 show the $F$ value and the associated probability for the \enhft.
Columns 4 and 5 indicate the same values for the \nest probe.
Columns 6 and 7, and 8 and 9, show the same values for the control star. 
The number of stars in the enhanced $F$ tests is one less than in the corresponding \nest probes because the brightest star in the field was used as reference.
The number of stars in the control tests is one less than in the target quasar because the star most similar in brightness to the quasar was used as control.

\subsubsection{Enhanced \ftest}

\begin{figure}[t!]
	\savebox{\IBox}{\includegraphics[trim = 0mm 12mm 0mm 20mm, clip, width=\linewidth]{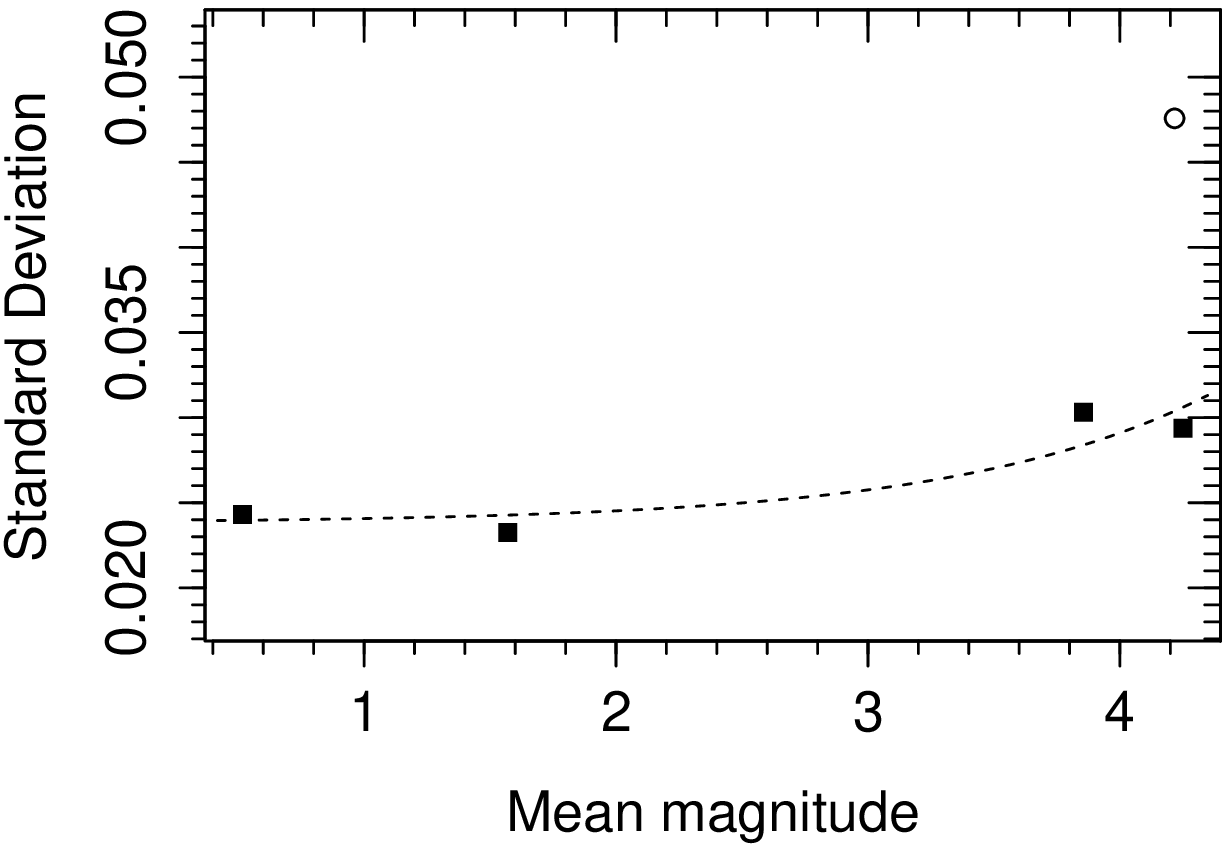}}
	\begin{pspicture}(\wd\IBox,\ht\IBox)%
	  \rput[lb](0,0){\usebox{\IBox}}%
	  \rput[lt](.2\wd\IBox,.85\ht\IBox){(a)} 
	\end{pspicture}
	\savebox{\IBox}{\includegraphics[trim = 0mm 0mm 0mm 20mm, clip, width=\linewidth]{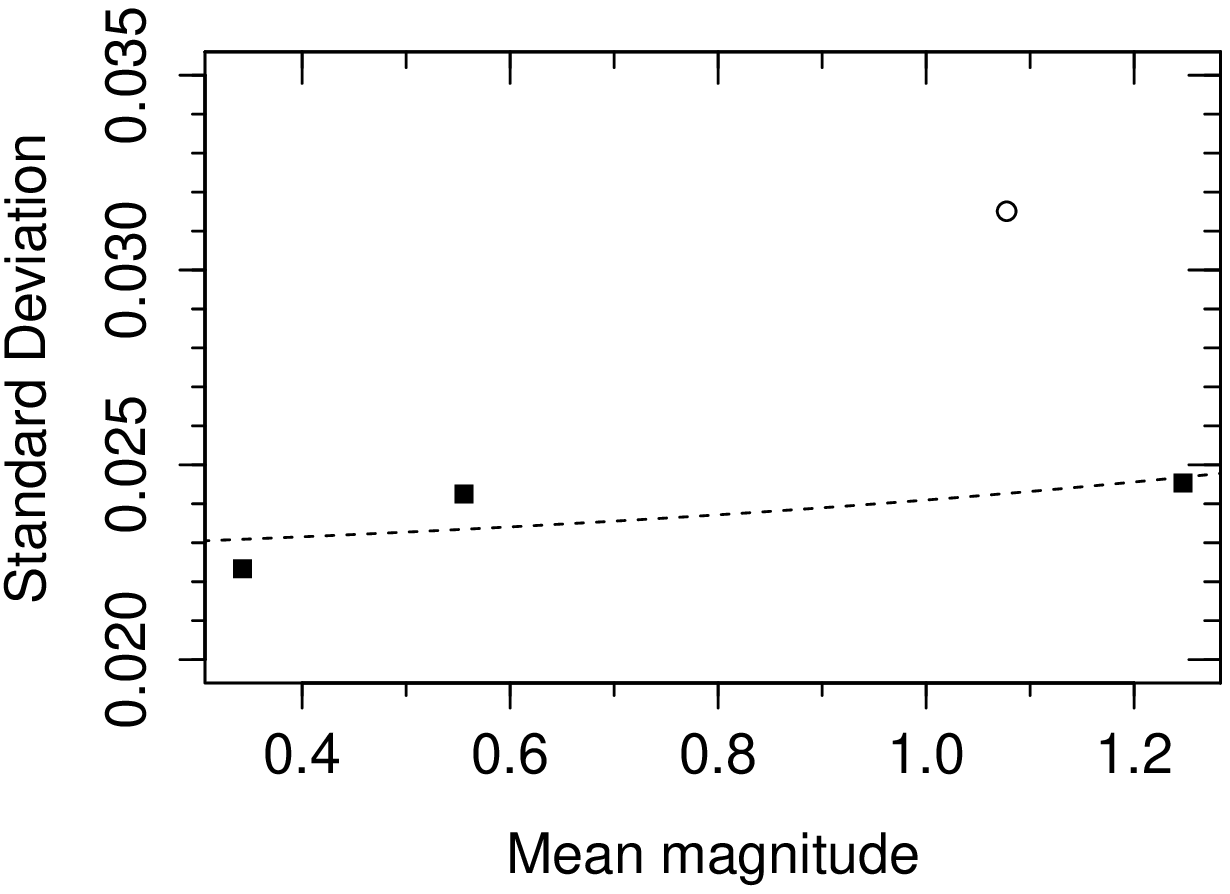}}
	\begin{pspicture}(\wd\IBox,\ht\IBox)%
	  \rput[lb](0,0){\usebox{\IBox}}%
	  \rput[lt](.2\wd\IBox,.85\ht\IBox){(b)} 
	\end{pspicture}
	    \caption{Light curve standard deviations vs mean differential magnitudes dependence.
    The standard deviation of the light curves yield error estimates.
    Panel (a) shows four comparison stars (squares) in the quasar \qsoA (circle) field that have been used to fit an exponential curve to the relationship between the standard deviation and the mean differential magnitude (dashed line).
    Panel (b) is like the previous panel but for three comparison stars (squares) in the field of the quasar \qsoB (circle).
    These figures show that the data dispersion for the quasar was much larger than the dispersion for their respective comparison stars, indicating possible variations.}\label{fig:errorfit}
\end{figure}

We have used the brightest star as reference to build the photometric differential light curves.
To perform the \enhft it is necessary that the errors of the comparison stars are transformed to the same level as the errors of the quasar.
In principle, any function that reasonably fits the differential light curve standard deviations and the mean magnitudes relationship can be used.
However, in another work (Polednikova et al., in preparation) we have fitted exponential curves to a large set of comparison stars in larger CCD fields, and find that such a function accurately describes this relationship.
The function is of the form:
\begin{equation}
    s_i = s_0 + A e^{\overline{m}_i},
\end{equation}
where $s_i$ is the light curve standard deviation and $\overline{m}_i$ is the light curve mean magnitude for the $i$th star, while $s_0$ and $A$ are the parameters to be fitted.
Notice that fitting two parameters to the comparison stars data reduces in two units the number of degrees of freedom for the comparison stars of the \enhft.
Figure~\ref{fig:errorfit} shows the relationship between differential light curves standard deviation for the comparison stars and the quasars (excluded from the fit), and their mean differential magnitudes.
The exponential curve fits the data for the comparison stars up to a precision of $\approx 0.001$\,mag.

\begin{table*}[ht]
\centering
\begin{center}
\caption{Test results for \qsoB data.\smallskip}\label{tbl:1e15498}
\begin{tabular}{ccc@{ }cccccc@{ }ccc}
\tableline\tableline \noalign{\smallskip}
 & \multicolumn{5}{c}{\qsoB} && \multicolumn{5}{c}{Control} \\
\cline{2-6} \cline{8-12} \noalign{\smallskip} 
 N. Stars & \multicolumn{2}{c}{\ftest} && \multicolumn{2}{c}{\aov} && \multicolumn{2}{c}{\ftest} && \multicolumn{2}{c}{\aov} \\
 \cline{2-3} \cline{5-6} \cline{8-9} \cline{11-12} \noalign{\smallskip} 
 & $F$ & $p$ && $F$ & $p$ && $F$ & $p$ && $F$ & $p$ \\
\tableline \noalign{\smallskip}
	1 & 3.098 & 1.2E-2 && 5.231 & 1.0E-2 && 1.252 & 0.32 && 0.361 & 0.78 \\ 
	2 & 2.384 & 1.2E-2 && 7.198 & 2.8E-3 && 0.963 & 0.52 && 0.467 & 0.71 \\ 
	3 & 2.477 & 4.6E-3 && 5.808 & 7.0E-3 &&  --   &  --  && 1.867 & 0.18 \\ 
	4 &  --   &   --   && 6.829 & 3.6E-3 &&  --   &  --  &&  --   &  --  \\ 
\tableline
\end{tabular} 
\tablecomments{The number of degrees of freedom for the \enhft is 19 for the numerator and 17, 36, and 55 for the denominator, for 1, 2, and 3 stars respectively.
For the \nest, the number of degrees of freedom is always 3 for the numerator and 16 for the denominator. }
\end{center}
\end{table*}

The data for the quasars shown in Figure~\ref{fig:errorfit} are clearly far away from stars of comparable brightness and the fitted exponential curve, indicating that their standard deviation may be dominated by variability rather than photometric errors.
Yet, the \ftest with a single comparison star fails to detect variations in \qsoA at the level of significance $\alpha = 0.01$, and 4 comparison stars are needed to reach $\alpha = 0.001$ (Table~\ref{tbl:us3150}).
In the case of \qsoB, the \enhft only reaches the level of significance $\alpha = 0.01$ when using three stars.
The $F$ statistics for the \enhft on both quasars shows small fluctuations but a monotonic decrease in the $p$ value.
Finally, the \enhft for the control stars also shows small fluctuations (in \qsoA) for both the $F$ statistics and the $p$ values, but in no case do the tests show any evidence of variation.

\subsubsection{Nested \aov}

Results of the \nest probes presented in Tables~\ref{tbl:us3150} and \ref{tbl:1e15498} show fluctuations for the $F$ statistics and the $p$ values for both the quasars and the control stars. 
In the case of \qsoA, all the tests yield  results below the significance level $\alpha = 0.001$, but for \qsoB the $p$ values are only below $\alpha = 0.01$.
In both cases, the control stars do not show any evidence of variations.

The results obtained by both the \enhft and \nest agree that \qsoA shows strong evidence of microvariability, while for \qsoB the evidence is not conclusive, as reported in \citet{diego:1998}.
They also agree with the results from the simulations, in the sense that \nest has more power to detect microvariations than the \enhft when there is a limited number of available field stars.



\section{Conclusions}


In this paper we have presented the \nest test and have compared the power of this probe with the \enhft presented in \citetalias{diego:2014}.
These tests use several field stars to analyze the light curve of a target quasar.
Both the \enhft and \nest are very powerful to detect photometric variations in differential light curves, and they asymptotically tend to larger power values than their respective single star counterparts (the \ftest and \aov).
The \nest probe shows a superior performance when the number of available bright stars is limited, but the \enhft surpasses \nest when there are several bright stars, some of them brighter than the target object.
This is a consequence of using a bright reference star by the \enhft that diminishes the errors of the differential photometry.

In our simulations, the power increases with the number of stars included in the analyses from approximately 40\% to 65\% for the \nest probes, and from 5\% to 90\% for the \enhft, both at the significance level of $\alpha = 0.001$.
These results agree with those obtained from the observed light curve of quasars \qsoA and \qsoB, for which the probabilities tend to lower values as the number of stars increases, while \nest yields lower probabilities than the \enhft due to the limited number of field stars. 
In this paper, we have demonstrated that the power of microvariability studies increases significantly by including several stars in the analysis, and that this procedure is an important improvement with respect to all the other tests that have been used in the quasar microvariability literature, from the \ctest to the single star \ftest and \aov, as well as multitesting techniques.

From these results, we conclude that both the \enhft and the \nest probe are robust techniques to detect quasar microvariability. 
They do not only add power to the microvariability detection, but by introducing several stars in the light curve analysis, ensure more reliable results than those obtained when using a single star.
In small fields with a limited number of stars, we encourage the use of the \nest rather than the \enhft because under these circumstances \nest is more powerful, and it can include one more bright star in the analysis.

The use of both the \enhft and \nest in future research will improve the detections of microvariability in quasars.
This improvement will eventually yield a better understanding of the physics involved in microvariability phenomena and the central quasar's engine.
The \enhft can be easily applicable to select varying sources candidates (i.e. quasars and variable stars) in regions visited several times during large photometric surveys, such as the the Legacy Survey Stripe 82 from the Sloan Digital Sky Survey.
The wide field images in such surveys, that include a large number of stars, allow to build a mean reference star with sources different to the stars used for comparison in the \enhft, thus ensuring the independence of the reference mean star and the comparison stars. 

\acknowledgements

This research has been supported by the UNAM-DGAPA -- PAPIIT IN110013 Program and the Canary Islands CIE: Tricontinental Atlantic Campus. This work is based upon observations acquired at the Observatorio Astron\'omico Nacional in the Sierra San Pedro M\'artir (OAN-SPM), Baja California, Mexico.
JP, AB, AMPG and JC are grateful for the support from the grant AYA2011-29517-C03-01 of the Spanish Ministerio de Econom\'ia y Competitividad (MINECO).
MADL wants to thank the UC MEXUS-CONACYT Postdoctoral Fellow in the UCR Department of Physics and Astronomy.
The authors are thankful to Jana Benda for the linguistic revision and to the anonymous referee for the constructive suggestions.

\bibliographystyle{apj}
\bibliography{jdo}

\appendix

\section{Nested ANOVA}\label{app:nesttheor}

The \nest analysis presented in this paper involves simulations of $a=7$ groups, each of them consisting of $b=5$ images (35 in total) of a target quasar and a number $n \leq 21$ of field stars that can be used for replicate measurements of the quasar differential photometry.
In practice, the images contained in a given group were obtained in a time lapse that is short (less than 10 min) in comparison with the microvariability time scale of the quasar, while the gap between groups of observations that will be compared is arbitrary.

This methodology implies $n$ \emph{pseudoreplicates} per image to yield a total of $35 n$ readings of the quasar differential photometry.
We use the term pseudoreplicates for the $n$ readings of a given image to note that these readings are not independent because we use the same quasar observation (the same image) to estimate the quasar differential photometry with $n$ stars. 
Images are the experimental units of our study; as we have 35 images (34 degrees of freedom in total), no matter how many stars we use in our analysis, the degrees of freedom are divided into 6 for the (7) groups of images and 28 ($=34-6$) for errors.
The number of replicate measurements that we perform on a given image will improve the precision on that image, and the variability among the images from a given group yields a measurement of the variability within that group.

\begin{figure*}[t]
\center
\includegraphics[angle=-90,trim=10.9cm 6cm 6.5cm 11cm,clip,width=\linewidth]{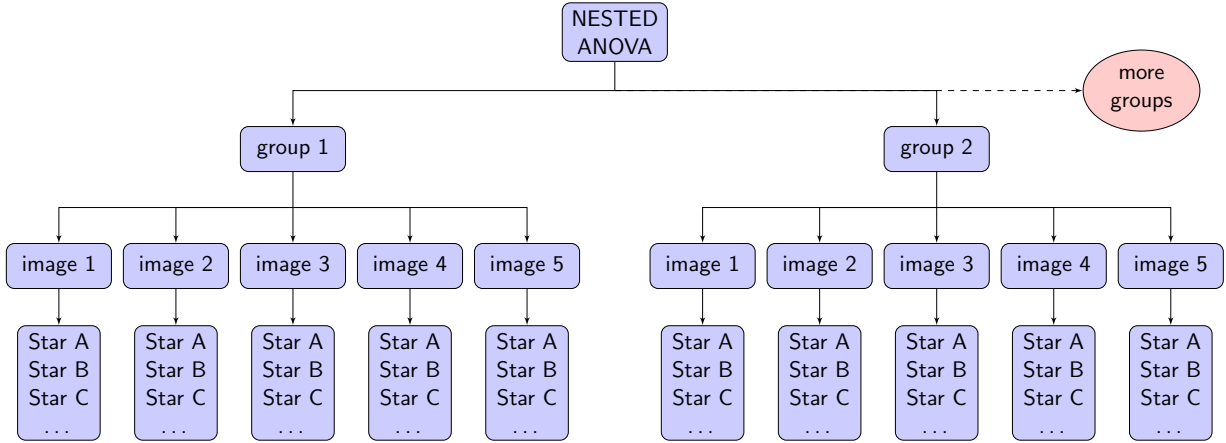}

\caption{\nest diagram. The goal is to test differences between groups of observations. Each group consists of 5 images that include the quasar and a number of stars. We can obtain a differential photometry measurement of the quasar from each star.}\label{fig:nestedaov}
\end{figure*}

Figure~\ref{fig:nestedaov} shows the schematic representation of the observational methodology: an arbitrary number of groups of observations (in our example 7), 5 images of the quasar in each group, and $n$ differential photometry measurements from each image, using $n$ different reference stars.
In fact, this is a two-stage nested design, probably the simplest \nest. Following \citet[chapter 14]{montgomery:2013}, the linear statistical model is expressed by:
\begin{equation}\label{eq:model}
    y_{ijk} = \mu + \gamma_i + \omega_{j(i)} + \varepsilon_{ijk},
\end{equation} 
where $y_{ijk}$ corresponds to the measurement of the quasar differential photometry using the reference star $k=1,2,\ldots,n$, located in the image $j=1,2,\ldots,b$ of the $i=1,2,\ldots,a$ group of observations. 
The true mean of the quasar light curve is denoted by $\mu$ and the deviation from the true mean of the $i$th group is denoted by $\gamma_i$. The deviation of the $j$th image of the quasar with respect to the mean of the $i$th group is denoted by $\omega_{j(i)}$, where the subscript $j(i)$ indicates that the $j$ observation is nested under the $i$th group.
Finally, $\varepsilon_{ijk}$ corresponds to the error term.

The model described by equation \ref{eq:model} yields a sum of (deviation) squares that, after some manipulation \citep{montgomery:2013}, can be expressed as:
\begin{equation}\label{eq:sssums}
    \sum_{i=1}^a \sum_{j=1}^b \sum_{k=1}^n (y_{ijk} - \bar{y})^2 = b n \sum_{i=1}^a (\bar{y}_i - \bar{y})^2 + n \sum_{i=1}^a \sum_{j=1}^b (\bar{y}_{ij} - \bar{y}_i)^2 + \sum_{i=1}^a \sum_{j=1}^b \sum_{k=1}^n (y_{ijk} - \bar{y}_{ij})^2,
\end{equation}
where the horizontal line over a letter indicates the mean value [e.g., $\bar{y}_i = (\sum_{j=1}^b \sum_{k=1}^{n} y_{ijk}) / bn$]. 
Equation~(\ref{eq:sssums} can also be expressed symbolically as:
\begin{equation}
    SS_T = SS_G + SS_{O(G)} + SS_E,
\end{equation}
where $SS_T$ denotes the total sum of squares, $SS_G$ the sum of squares due to groups, $SS_{O(G)}$ the sum of squares due to the nested observations in groups, and $SS_E$ the residual sum of squares due to errors.
The degrees of freedom for each square sum is: $a b n - 1$ for $SS_T$, $a - 1$ for $SS_G$, $a (b-1)$ for $SS_{O(G)}$, and $a b (n - 1)$ for $SS_E$.
Dividing a sum of squares $SS$ by the respective degrees of freedom $\nu$ yields the mean square ($MS = SS / \nu$).
The ratios between mean squares are distributed as $F$.



In our simulations, the groups of observations are drawn from an infinite pool of instants in the continuous light curve of a target object, and the observations in each group are drawn from an infinite pool of possible images, thus both groups and observations constitute random effects.
In this case we assume that $\gamma_{i}$ is NID(0,$\sigma^{2}_{\gamma}$) and $\omega_{j(i)}$ is NID(0,$\sigma^2_{\omega}$), where NID(0,$\sigma^2$) means normally and independently distributed with mean 0 and variance $\sigma^{2}$. 
Then we test $H_{0}:\sigma_{\gamma}^{2}\;=\;0$ by $MS_{G}/MS_{O(G)}$ and $H_{0}:\sigma_{\omega}^{2}\;=\;0$ by $MS_{O(G)}/MS_{E}$. 

The residuals for a two-stage nested design are given by
\begin{equation}
e_{ijk} = y_{ijk} - \bar{y}_{ij},
\end{equation}
where $\bar{y}_{ij}$ are the individual image averages, and the fitted value is:
\begin{equation}
\hat{y}_{ijk} = \bar{y}+\hat{\gamma}_{i}+\hat{\omega}_{j(i)}
\end{equation}

For random effects, the analysis of variance method allows to estimate the variance components $\sigma^{2}$, $\sigma_{\beta}^{2}$, and $\sigma_{\tau}^{2}$ using the expected mean squares:

\begin{equation}
\hat{\sigma}^{2} = MS_{E}
\end{equation}
\begin{equation}
\hat{\sigma}^{2}_{\beta} = \frac{M\mbox{S}_{O\left( G \right)}-M\mbox{S}_{E}}{n}
\end{equation}
\begin{equation}
\hat{\sigma}^{2}_{\tau} = \frac{M\mbox{S}_{G}\; -\; M\mbox{S}_{O\left( G \right)}}{bn}
\end{equation}

\section{Nested Analysis of Variance R code}\label{app:nestcode}

\begin{verbatim}
# Nested ANOVA
# (c) J. A. de Diego - November 2014 

# ****************************
# Input file example (file without headers)
# TIME        QUASAR    S1        S2        S3        S4        S5        GROUP
# 03:08:50    21.504    17.389    17.851    18.911    21.187    21.577    1
# 03:10:44    21.557    17.347    17.882    18.951    21.232    21.655    1
# 03:12:34    21.488    17.400    17.840    18.913    21.207    21.597    1
# 03:14:24    21.481    17.319    17.844    18.908    21.178    21.593    1
# 03:15:54    21.484    17.348    17.841    18.919    21.176    21.577    1
# 04:31:24    21.601    17.354    17.877    18.938    21.232    21.611    2
# 04:33:29    21.552    17.338    17.872    18.930    21.222    21.597    2
# 04:34:54    21.635    17.368    17.895    18.966    21.217    21.647    2
# 04:36:19    21.583    17.334    17.863    18.925    21.233    21.607    2
# 04:37:44    21.581    17.363    17.894    18.963    21.240    21.651    2

# The number of columns for the stars is arbitrary

# Magnitudes for the quasar and stars are as generated byt IRAF
# routines, i.e. unprocessed (not differential or extinction corrected)

# Columns TIME and GROUP are not used in this code, but probably needed 
# for figures and other computations.

# This code assumes that the last star corresponds to the control.

# ****************************
# Output
#   nest.anova.sum.qv   Contains a list with the results of tests for the quasar
#   nest.anova.sum.qn   Contains a list with the results of tests for the control star
#   prob.anova.qv       Is a vector with the probabilities of tests for the quasar
#   prob.anova.qn       Is a vector with the probabilities of tests for the control star
# ****************************


# **********************
# Read data from "mydata.dat" and prepare variables
(test <- read.table("US3150_V3.dat"))
s <- length(test) -2    # Number of sources in the field (including the quasar)
n <- dim(test)[1]       # Data points in the light curve
ng <- 5                 # Number of points in each group
g <- n/ng               # Number of groups

qlc <- test
colnames(qlc) <- c("Time","q",paste0("s",seq(1:(s-1))),"Group")

dqlc <- data.frame(
  qv =  as.vector(as.matrix(qlc$q-qlc[,c(paste0("s",seq(1:(s-1))))])),  # Variable quasar
  qn =  as.vector(as.matrix(qlc$s5-qlc[,c(paste0("s",seq(1:(s-1))))])),  # Control star
  st = rep(1:(s-1), each=n),        # Star factor
  ag = rep(1:g, each=ng),           # ANOVA group factor (or read column Group)
  ao = rep(1:ng)                    # ANOVA group element factor
)

# ****************************

# Allocate matrices
nest.anova.sum.qv <-vector("list")
nest.anova.sum.qn <-vector("list")
prob.anova.qv <- matrix(,nrow = 1, ncol = (s-1))
  colnames(prob.anova.qv) <- paste('cs.',1:(s-1), sep="")
prob.anova.qn <- matrix(,nrow = 1, ncol = (s-1))
  colnames(prob.anova.qn) <- paste('cs.',1:(s-1), sep="")

# Perform test
for(j in 2:(s)) {
  # Temporal quasar variables
  dlcj <- dqlc[dqlc$st < j,] 
  ym <- as.vector(tapply(dlcj$qv,
                         list(dlcj$ag, dlcj$ao), 
                         mean)) # vector of observation (image) means
  tm <- factor(as.vector(tapply(as.numeric(dlcj$ag),                               
                                list(dlcj$ag, dlcj$ao), 
                                mean))) # group factors
  # Test and probabilities for the quasar
  nest.anova.sum.qv[[(j-1)]] <- summary(aov(ym~tm))
  prob.anova.qv[j-1] <- nest.anova.sum.qv[(j-1)][[1]][[1]][,"Pr(>F)"][[1]]
  # Temporal control variables
  ym <- as.vector(tapply(dlcj$qn,
                         list(dlcj$ag, dlcj$ao), 
                         mean)) # vector of obsevation (image) means
  tm <- factor(as.vector(tapply(as.numeric(dlcj$ag),                               
                                list(dlcj$ag, dlcj$ao), 
                                mean))) # group factors
  # Test and probabilities for the quasar
  nest.anova.sum.qn[[(j-1)]] <- summary(aov(ym~tm))
  prob.anova.qn[j-1] <- nest.anova.sum.qn[(j-1)][[1]][[1]][,"Pr(>F)"][[1]]
}

# Remove last element for the control star results (it is a test with itself)
nest.anova.sum.qn[[(j-1)]] <- NULL
prob.anova.qn <-  prob.anova.qn[-(j-1)]

\end{verbatim}

\end{document}